\newcommand{\msol}{\mbox{\,M$_\odot$}}
\newcommand{\mdotrate}{\mbox{\,M$_\odot$\,yr$^{-1}$}}
\documentclass[a4,useAMS,usenatbib]{mn2e}

\usepackage{graphicx}

\title[Massive star formation simulation algorithms]{Radiation-hydrodynamical simulations of massive star formation using Monte Carlo radiative transfer \\ I. Algorithms and numerical methods}
\author[Tim J. Harries]{Tim J. Harries\thanks{E-mail:
th@astro.ex.ac.uk} \\
Department of Physics and Astronomy, University of Exeter, Stocker Road, Exeter EX4 4QL, United Kingdom\\}
\begin{document}

\date{}

\pagerange{\pageref{firstpage}--\pageref{lastpage}} \pubyear{2015}

\maketitle

\label{firstpage}

\begin{abstract}
We present a set of new numerical methods that are relevant to calculating radiation pressure terms in hydrodynamics calculations, with a particular focus on massive star formation. The radiation force is determined from a Monte Carlo estimator and enables a complete treatment of the detailed microphysics, including polychromatic radiation and anisotropic scattering, in both the free-streaming and optically-thick limits. Since the new method is computationally demanding we have developed two new methods that speed up the algorithm. The first is a photon packet splitting algorithm that enables efficient treatment of the Monte Carlo process in very optically thick regions. The second is a parallelisation method that distributes the Monte Carlo workload over many instances of the hydrodynamic domain, resulting in excellent scaling of the radiation step.  We also describe the implementation of a  sink particle method that enables us to follow the accretion onto, and the growth of, the protostars. We detail the results of extensive testing and benchmarking of the new algorithms.
\end{abstract}


\section{Introduction}

Massive stars are hugely important in galactic ecology, enriching them
chemically and providing strong feedback effects via radiation,
stellar winds and supernovae. They also play a pivotal role in
measuring star formation rates in distant galaxies, and therefore
determining the star formation history of the Universe
\citep{kennicutt_1998}. However, the process by which massive stars
form is less well understood than that of lower-mass stars, both
observationally, because massive protostars are rare and difficult to
observe, and theoretically because radiation feedback has a much
stronger influence on the gas dynamics than it does for solar-mass
objects. Broadly it appears that observational and theoretical
perspectives on massive star formation are converging towards a
scenario in which a massive young stellar object contains a central
($>10$\,M$_\odot$) protostar surrounded by a stochastically accreting
Keplerian disc and a stellar- or disc-wind outflow, although
substantial conflicts between the predictions of numerical models
remain \citep{tan_2014}.

It has been known for some time that massive protostars undergoing
spherical accretion will eventually produce sufficient radiation
pressure to halt the inflow of material
(e.g. \citealt{wolfire_1987}). Extending models to multiple
dimensions, and the introduction of angular momentum, leads to a
scenario in which the protostar can accrete via a disc, which
intercepts a smaller fraction of the protostellar radiation field, and
the radiation escapes preferentially via low-density cavities along
the rotation axis (e.g. \citealt{yorke_2002}). Recent work by
\cite{krumholz_2009} used 3-D adaptive mesh refinement (AMR)
simulations coupled with a flux-limited diffusion (FLD) approximation
for the radiation transport (RT) to simulate the formation of a
massive binary. They found accretion occurred via a radiatively-driven
Rayleigh-Taylor instability that allowed material to penetrate the
radiation-driven cavities above and below the protostars. Their
simulation resulted in the formation of a massive (41\msol and
29\msol) binary pair.

\cite{kuiper_2010a} implemented a hybrid RT scheme in which ray
tracing is used to calculate the radiation field from the protostellar
photosphere out to where the material becomes optically thick, at
which point the ray-traced field becomes a source term in the grey
diffusion approximation used for the rest of the domain. The method
shows good agreement with more detailed treatments of the RT but has
vastly reduced computational cost
(\citealt{kuiper_2013}). Two-dimensional calculations adopting this
method resulted in the formation of massive stars via stochastic disc
accretion accompanied by strong radiatively driven outflows
(\citealt{kuiper_2010b}).

Radiatively-driven Rayleigh-Taylor instabilities have not been
identified in the hybrid RT models (\citealt{kuiper_2012}), suggesting
a fundamental difference with the Krumholz et al. computations. It
appears that the grey FLD method considerably underestimates the
driving force of the radiation, since the opacity used for the
radiation pressure calculation is the Rosseland opacity at the dust
temperature, whereas in the hybrid case the primary momentum
deposition occurs where the radiation from the protostar is incident
on the dust (near the dust sublimation radius). At this point the
radiation temperature is much higher, and the dust is much more
opaque. It is possible that the grey FLD method underestimates the the
driving force by up to two orders of magnitude
(\citealt{kuiper_2012}).

The fundamental point here is that the dynamics of the simulations can
be critically affected by the level of microphysical detail
employed. The lessons of recent numerical simulations are that disc
accretion may occur, and fast, wide-angle outflows may be driven. It
seems that radiation pressure does not present a fundamental physical
barrier to massive star formation via accretion, although the
strength, geometry, and longevity of the outflows is still
contentious.  Furthermore, neither the \cite{krumholz_2009} models nor
those of the \cite{kuiper_2010b} include the effects of
photoionization, which will come into play as the protostar shrinks
towards the main sequence and its radiation field hardens. All these
factors point towards the necessity for a more thorough treatment of
the microphysical processes that underpin that radiation feedback,
such as a polychromatic prescription for the radiation field,
consideration of both dust and gas opacity (absorption and
scattering), and the inclusion of photoionization.

\textbf{Recently significant progress has been made in using MC
  transport in RHD codes. For example \cite{nayakshin_2009} used MC
  photon packets to treat radiation pressure in smoothed-particle
  hydrodynamics. \cite{acreman_2010} presented a proof-of-concept
  calculation involving combining 3D smooth-particle hydrodynamics
  with MC radiative equilibrium. In this case the SPH particle
  distribution was used to construct an AMR grid on which the MC
  radiative-equilibrium calculation was conducted. We then developed
  an Eulerian hydrodynamics module to perform 3D hydrodynamics on the
  native AMR grid used for the radiation transfer
  \citep{haworth_2012a}. A similar approach was adopted by
  \cite{noebauer_2012}, who incorporated a radiation-pressure force
  estimator as well as a radiative-equilibrium
  calculation. \cite{roth_2014} also coupled MC transport and 1D
  hydrodynamics, incorporating special relativity and resonance line
  scattering.}
 
We have now developed an AMR RHD code ({\sc torus}) that employs a
highly-parallelised, time-dependent, Monte-Carlo (MC) method
\citep{harries_2011} to follow the RT at a level of microphysical
detail comparable to that of dedicated RT codes such as Cloudy
\citep{ferland_2013}.  The consequences of this breakthrough are
two-fold: firstly, the increased sophistication of the treatment of
the radiation field can have a significant impact on the hydrodynamics
\citep{haworth_2012a}, and secondly it becomes possible to make a much
more direct comparison between the models and observations across a
wide range of continuum (e.g. near-IR dust, radio free-free) and line
(e.g. forbidden, molecular or recombination) diagnostics,
\citep{rundle_2010, haworth_2012b, haworth_2013}.

In this paper we describe the implementation of (i) a new, highly
parallelised method for calculating the radiation pressure using Monte
Carlo estimators, (ii) a packet splitting method that improves the
efficiency of the method at high optical depths, and (iii) a
Lagrangian sink-particle algorithm that enables us to follow the
growth of the protostar. We provide the results of extensive
benchmarking of the new methods, with a view to conducting simulations
of massive star formation that incorporate both radiation pressure and
ionisation feedback.

\section{Radiation pressure}

Our Monte Carlo method is closely based on that of \cite{lucy_1999},
in which the luminosity $L$ of the illuminating object is divided up
into $N$ discrete photon packets that have a constant energy as they
propagate through the adaptive mesh. The energy of each packet is
\begin{equation}
\epsilon_i = w_i L \Delta t
\end{equation}
where $\Delta t$ is the duration of the Monte Carlo experiment and
$w_i$ is a weighting factor, normalized so that
\begin{equation}
\sum_{i=1}^{N} w_i = 1.
\end{equation}
Note that the original Lucy method has $w_i = 1/N$, so that each
photon packet had the same energy.

As the packets propagate through the grid they may undergo absorption
or scattering events. During an absorption event the photon packet is
immediately re-emitted, with a new frequency selected at random from a
probability density function created from the appropriate emissivity
spectrum at that point in the grid. The radiation field therefore
remains divergence free during the calculation. After the $N$ packets
have been propagated, the energy density for each cell can be
straightforwardly computed and hence the absorption rate can be
determined.

The calculation of the radiation pressure may be conveniently
conducted in parallel with the radiative equilibrium calculation. The
simplest method is to assign a momentum change to each cell ($\Delta
{\bf p_j}$), which is zeroed at the start of each MC loop. The each
time a photon packet enters and then leaves a cell the difference in
momentum between the incoming and outgoing momenta is added to the
cell's momentum change
\begin{equation}
\Delta {\bf p_j} = \Delta {\bf p_j} + \frac{\epsilon_i}{c} \left( {\bf {\hat u}_{\rm in}} - {\bf {\hat u}_{\rm out}} \right)
\end{equation}
where ${\bf {\hat u}_{\rm in}}$ is the direction vector of the photon
packet when it enters the cell and ${\bf {\hat u}_{\rm out}}$ is the
direction vector of the photon packet when it exits the cell. At the
end of the MC loop we can then calculate the radiation-force per unit
volume for each cell via
\begin{equation}
{\bf f}_{{\rm rad},j} = \frac{\Delta {\bf p_j}} {\Delta t V_j}
\end{equation}
where $V_j$ is the cell volume. We will call this the momentum tracking algorithm.

This method has the advantage that it is straightforward to implement,
is transparently related to the underlying physics, is valid for
arbitrary scattering phase matrices, and is very fast to calculate. It
does however have the disadvantage that the number of
scatterings/absorptions per cell tends to zero in the optically thin
limit. This is an analogous problem to that associated with the
\cite{bjorkman_2001} radiative-equilibrium method.

An alternative method is to use a MC estimator of the radiation vector
flux to calculate the radiation force. We define the energy density
$u_\nu$ as the energy density per unit frequency of a beam of
radiation. The energy carried in a beam of volume $dV$ is
\begin{equation}
dE = u_\nu dV d\nu 
\end{equation}
and since $dV = cdAdt$ we get
\begin{equation}
dE = u_\nu cdAdtd\nu  = I_\nu dt d\nu dA d\Omega
\end{equation}
where $I_\nu$ is the specific intensity. Hence we have
\begin{equation}
u_\nu = \frac{I_\nu}{c} d\Omega.
\label{u_i_rel}
\end{equation}
If a photon packet traverses length $\ell$ between successive events
(events being absorptions/scatterings or crossing cell boundaries)
then the energy density due to this part of the path of packet
$\epsilon_i$ is $(\epsilon_i / V_j) \delta t / \Delta t$ where $\delta
t = \ell / c$. We can therefore obtain an MC estimator of the energy
density
\begin{equation}
u_\nu d\nu = \frac{\epsilon_i \ell}{V_j c \Delta t}
\end{equation}
This may be related to the specific intensity via equation~\ref{u_i_rel} and we get
\begin{equation}
I_\nu d\Omega d\nu = \frac{\epsilon_i \ell}{\Delta t V_j}.
\end{equation}
The radiation flux vector of cell $j$ may then be written
\begin{equation}
{\bf F}_{j, \nu} = \int I_\nu (\Omega) {\bf d\Omega} = \frac{1}{\Delta t} \frac{1}{V_j} \frac{1}{d\nu} \sum_{d\nu} \epsilon_i \ell {\bf \hat{u}}
\end{equation}
where ${\bf \hat{u}}$ is the unit vector in the direction of the
photon packet as it traverses $\ell$ and only the paths for photon packets with frequencies
in the range ($\nu$, $\nu + d\nu$) are considered. The force per unit
volume for cell $j$ is then
\begin{equation}
{\bf f}_{{\rm rad},j} = \frac{1}{c} \int \kappa_\nu \rho {\bf F}_\nu
d\nu = \frac{1}{c}\frac{1}{\Delta t} \frac{1}{V_j} \sum \epsilon_i
\kappa_\nu \rho \ell {\bf \hat{u}}.
\end{equation}
where $\kappa_\nu$ is the opacity at the frequency $\nu$ of packet
$\epsilon_i$ as it traverses length $\ell$, $\rho$ is the mass
density, and the summation is now over all packets. Note that the
opacity $\kappa_\nu$ must consist of both the absorption and
scattering opacities
\begin{equation}
\kappa_i = (1-\alpha_\nu)\kappa_\nu + (1-g_\nu) \alpha_\nu \kappa_\nu
\end{equation}
where $\alpha_\nu$ is the albedo, and $g_\nu = \langle \cos \theta
\rangle$ where $\theta$ is the scattering angle (i.e. $g_\nu = 0$ for
isotropic scattering, positive for preferentially forward scattering
and negative for back scattering). Although marginally more expensive
to calculate, this flux estimator method should be superior in the
optically-thin limit, since even photon packets that do not interact
in a cell contribute to the estimator.

\subsection{Radiation pressure tests}

We constructed a suite of tests of our radiation pressure algorithms
based on a uniform-density sphere of radius $R_s = 0.1$\,pc and mass
100\msol\ illuminated by a solar-type star. The grey opacity $\kappa$
was selected in order that $\tau = \kappa \rho R_s = 0.1, 1, 10$ or
100. The models were run with a 1-D spherical geometry on a uniform
radial mesh comprising 1024 cells. We ran models for pure absorption
($\alpha = 0$) and pure scattering ($\alpha=1$) cases, and for the
scattering models we further subdivided the models into isotropic
scattering and forward scattering (using a Henyey-Greenstein phase
function with $g = 0.9$). We used $10^5$ photon packets for all the
models. For each model we calculated the radiation force per unit
volume ($F \kappa \rho /c$) for both the Monte Carlo flux estimator
and the momentum tracking algorithm.

\subsubsection{Pure absorption case}

In this case the radiation force per unit volume ($f_{\rm rad}$) as a
function of radius $r$ in the sphere is expected to be
\begin{equation}
f_{\rm rad} = \frac{L \kappa \rho \exp (-\tau(r))}{4 \pi r^2 c}
\end{equation}
where $\tau(r)$ is the radial optical depth from the centre of the
grid to $r$. Since the photon packets stop propagating when they are
absorbed the number of packets passing each cell will be monotonically
declining as a function of radius, and we therefore expect the noise
on the estimates of the radiation force to increase
radially. Furthermore we expect that the noise in the momentum
tracking algorithm to be larger than that of the flux estimator
method, particularly for the optically thin cases ($\tau = 0.1, 1$).

\begin{figure}
\includegraphics[width=80mm]{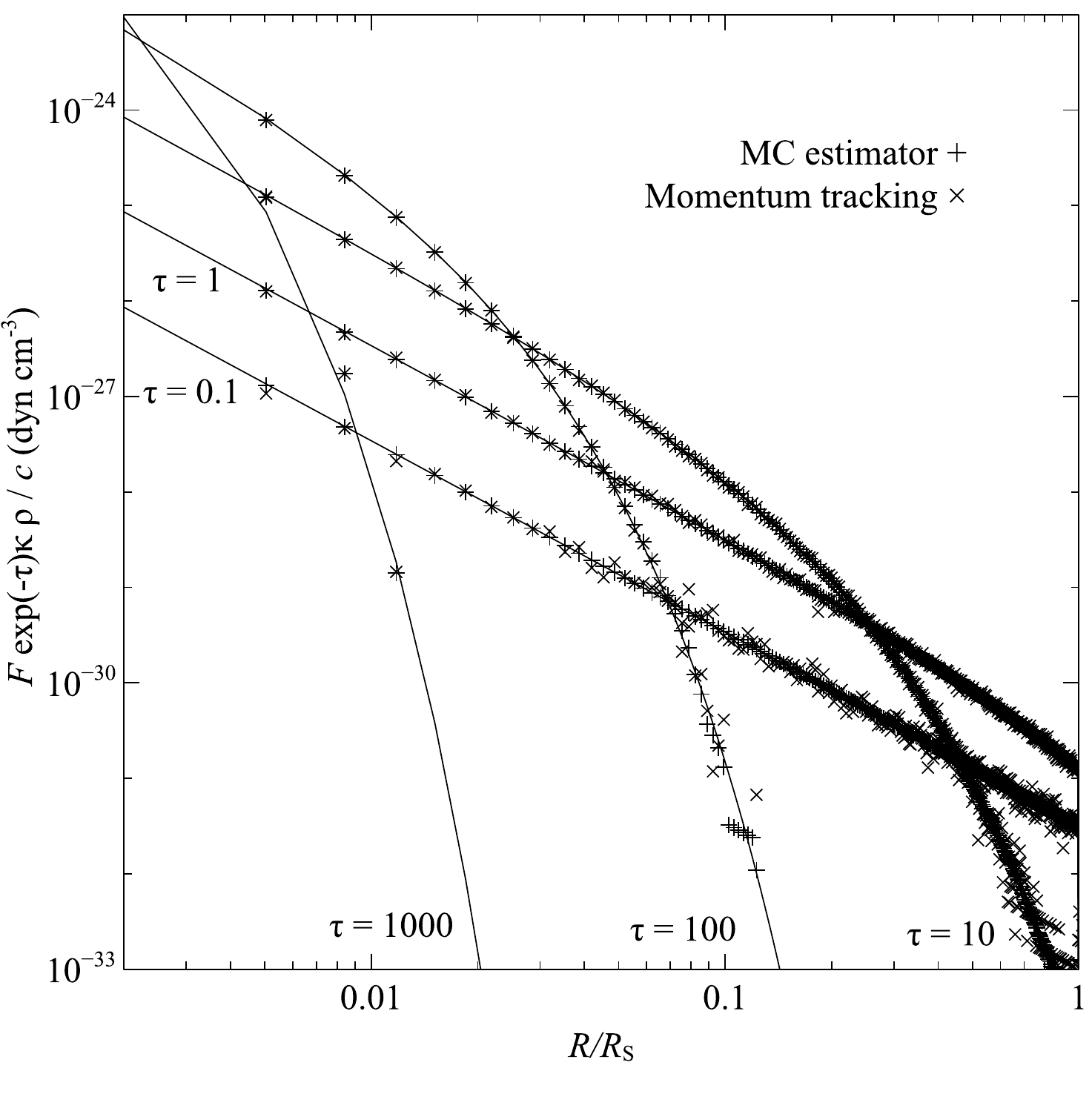}
\caption{Benchmark for absorption ($\alpha = 0$). The solid lines
  indicate the analytical results ($F \kappa \rho /c$), while values
  calculated from the Monte Carlo flux estimator are shown as plusses
  ($+$) and those found from following packet momenta are shown as
  crosses ($\times$). Benchmarks are plotted for spheres of four
  optical depths ($\tau = 0.1, 1, 10, 100$). }
\label{shelltest_abs_fig}
\end{figure}

The results of this benchmark are displayed in
Figure~\ref{shelltest_abs_fig} and the Monte Carlo estimators show
good agreement with the analytical result, with the expected noise
dependencies. These models ran rapidly, and since each photon packet
only undergoes one absorption event the higher optical depth cases ran
the most quickly.

\subsubsection{Pure scattering cases}

For isotropic scattering the number of photon interactions before
escape from an optically-thick, uniform spherical of radial optical
depth $\tau_{\rm max}$ will be
\begin{equation}
N_{\rm scat}  \approx \tau_{\rm max}^2 / \langle \tau^2 \rangle = \tau_{\rm max}^2 / 2.
\end{equation}
Thus for the $\tau =100$ model each photon packet will undergo $\sim
5000$ scattering events before escaping. Since the radiation field is
divergence free, the outward flux at every radius is a constant $F = L
/ (4 \pi r^2)$ and the radiation force per unit volume is then
\begin{equation}
f_{\rm rad} = \frac{L \kappa \rho}{4 \pi r^2 c}.
\end{equation}
Once again we expect the flux estimator method to have less noise than
the momentum tracking method, and that the latter algorithm will be
rather poorer in the low optical depth models.

\begin{figure}
\includegraphics[width=80mm]{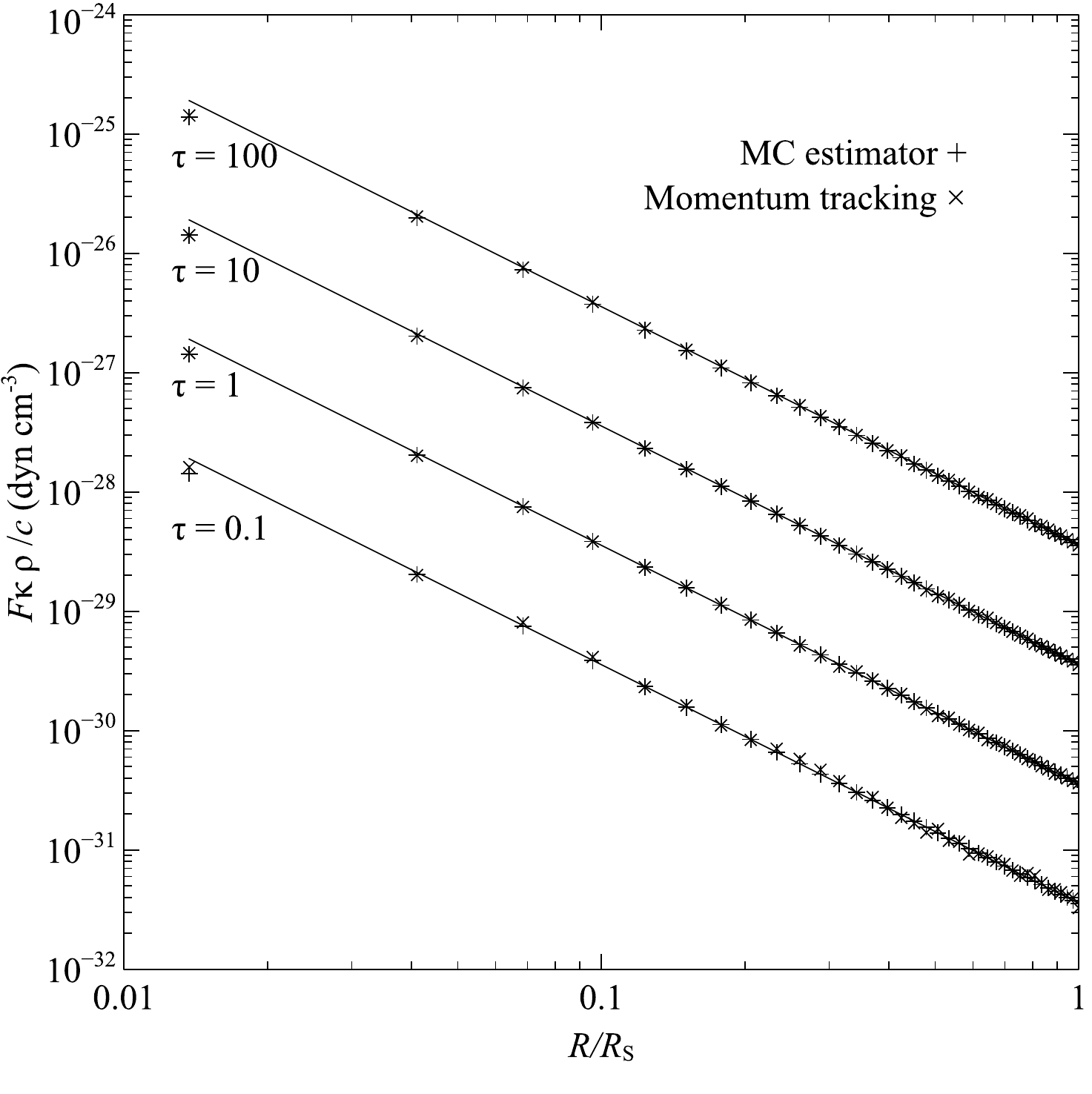}
\caption{Benchmarks for an isotropic scattering medium ($\alpha = 1$). Symbols are the same as those for Figure~\ref{shelltest_abs_fig}.}
\label{shelltest_sca_fig}
\end{figure}

The results of the isotropic, pure scattering benchmark are displayed
in Figure~\ref{shelltest_sca_fig}. Excellent agreement is seen between
the Monte Carlo estimators and the analytical result, and as expected
the noisiest estimator is the momentum tracking algorithm for the
$\tau=0.1$ model.

\begin{figure}
\includegraphics[width=80mm]{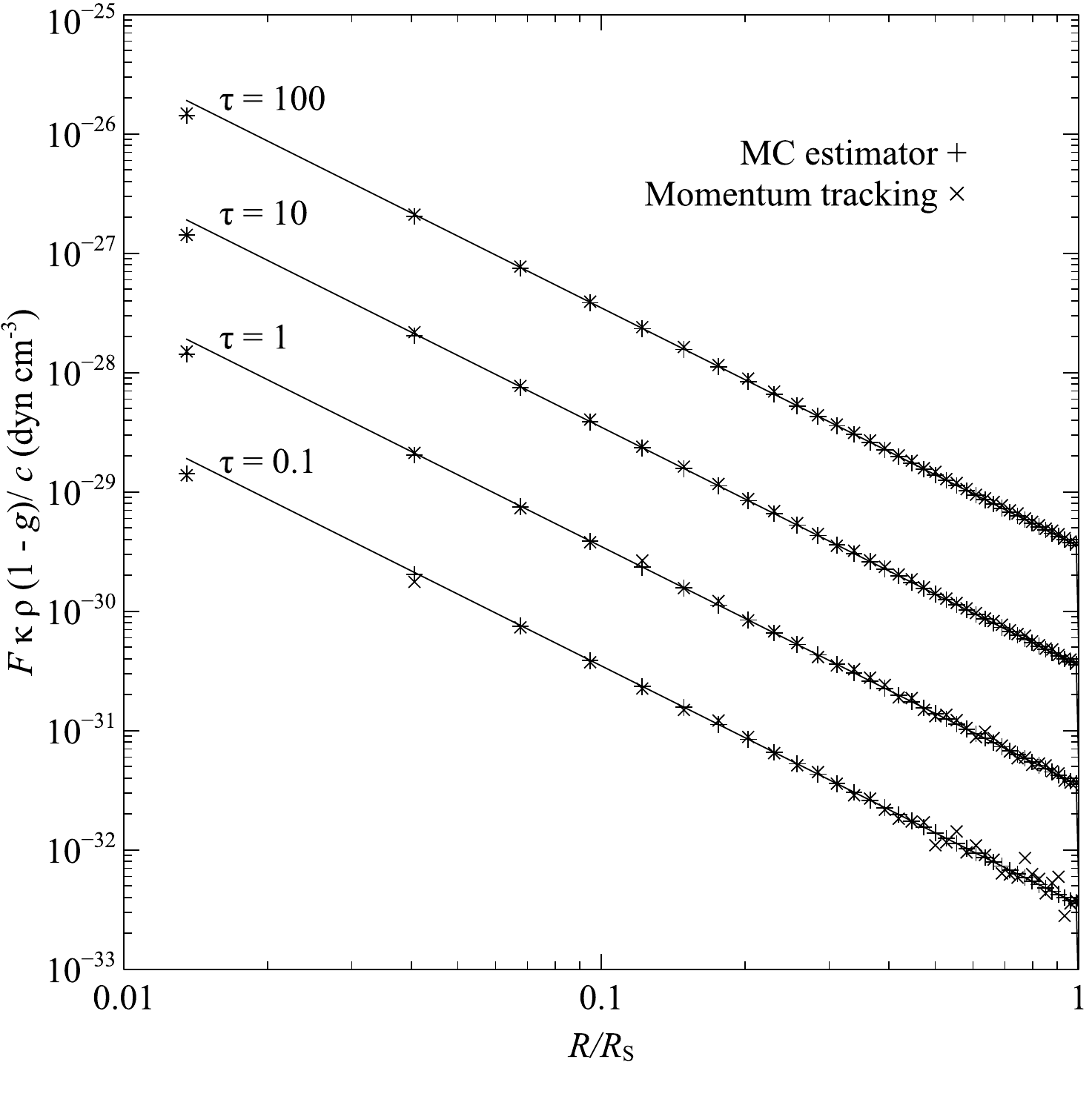}
\caption{Benchmarks for a scattering medium ($\alpha = 1$) with a forward-scattering Henyey-Greenstein phase function ($g = 0.9$). Symbols are the same as those for Figure~\ref{shelltest_abs_fig}.}
\label{shelltest_sca_9_fig}
\end{figure}

Finally we ran the pure scattering model for a forward-scattering Henyey-Greenstein phase function ($g = 0.9$). In this model the nett radiation force should be reduced by a factor of $(1-g)$ over the corresponding isotropic scattering case, i.e.
\begin{equation}
f_{\rm rad} = \frac{L \kappa \rho (1 - g)}{4 \pi r^2 c}.
\end{equation}
The results of the forward-scattering model are given in Figure~\ref{shelltest_sca_9_fig} and excellent agreement with the expected analytical radiation force is found.

\subsection{Radiation pressure hydrodynamic test}

Satisfied that the radiation pressure forces were being properly captured by the code, we progressed to testing the treatment in a dynamic model using a similar benchmark to that presented by \cite{nayakshin_2009}. A uniform density sphere ($\rho_0 = 1.67 \times 10^{-22}$\,g\,cc$^{-1}$ and $R_s = 10^{17}$\,cm) was modelled using 1-d grid comprising 1024 uniformly spaced radial cells. The sphere was illuminated by a 1\,$L_\odot$ star placed at the centre, and we used $10^5$ photon packets per radiation step.

For the purposes of testing we assumed isothermal gas at a temperature of 10\,K and considered cells above a critical density threshold ($1.67 \times 10^{-25}$\,g\,cc$^{-1}$) to be completely optically thick at all wavelengths with an albedo of zero, while cells with densities below the threshold were considered to be transparent to all radiation. Photon packets entering optically thick cells were immediately absorbed and were not remitted, allowing straightforward comparison with analytical results.

In this case the radiation pressure sweeps up a thin shell whose equation of motion (assuming a negligible contribution from thermal gas pressure) is given by
\begin{equation}
\frac{d}{dt} \left[ \left( \frac{4}{3}\pi R^3 \rho_0 \right) \dot{R}\right] = \frac{L}{c}.
\end{equation}
Integrating the above expression we get
\begin{equation}
R = \left( \frac{3 L}{2 \pi c \rho_0} \right)^{1/4} t^{1/2}.
\label{shell_eq}
\end{equation}

\begin{figure}
\includegraphics[width=80mm]{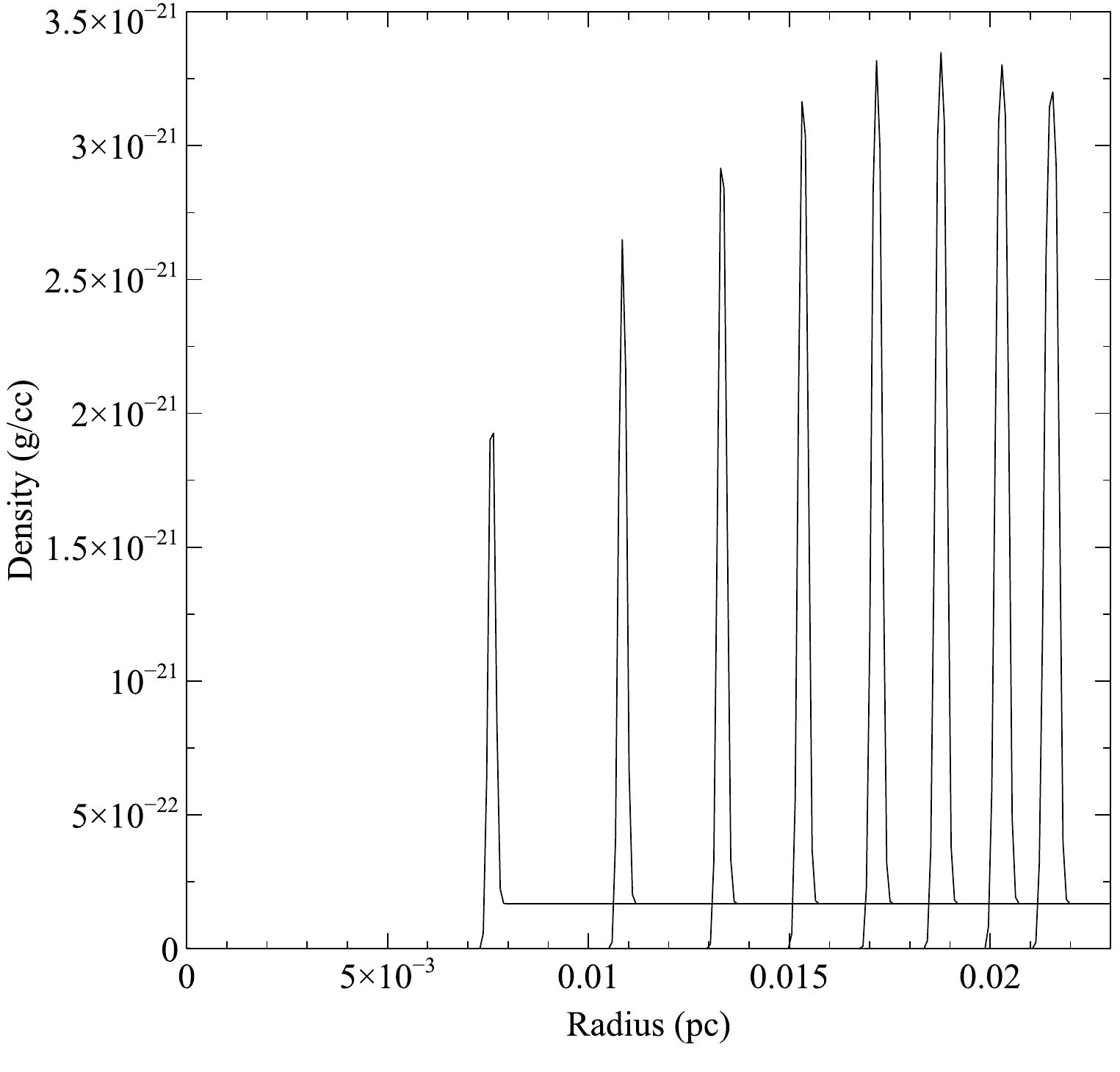}
\caption{The development of the radiation-pressure driven spherical shell as a function of time. The mass density as a function of radius is plotted at 1000\,kyr intervals from a simulation time of 1000\,kyr to 8000\,kyr. Some broadening of the shell is evident at later times.}
\label{shell_progression_fig}
\end{figure}

\begin{figure}
\includegraphics[width=80mm]{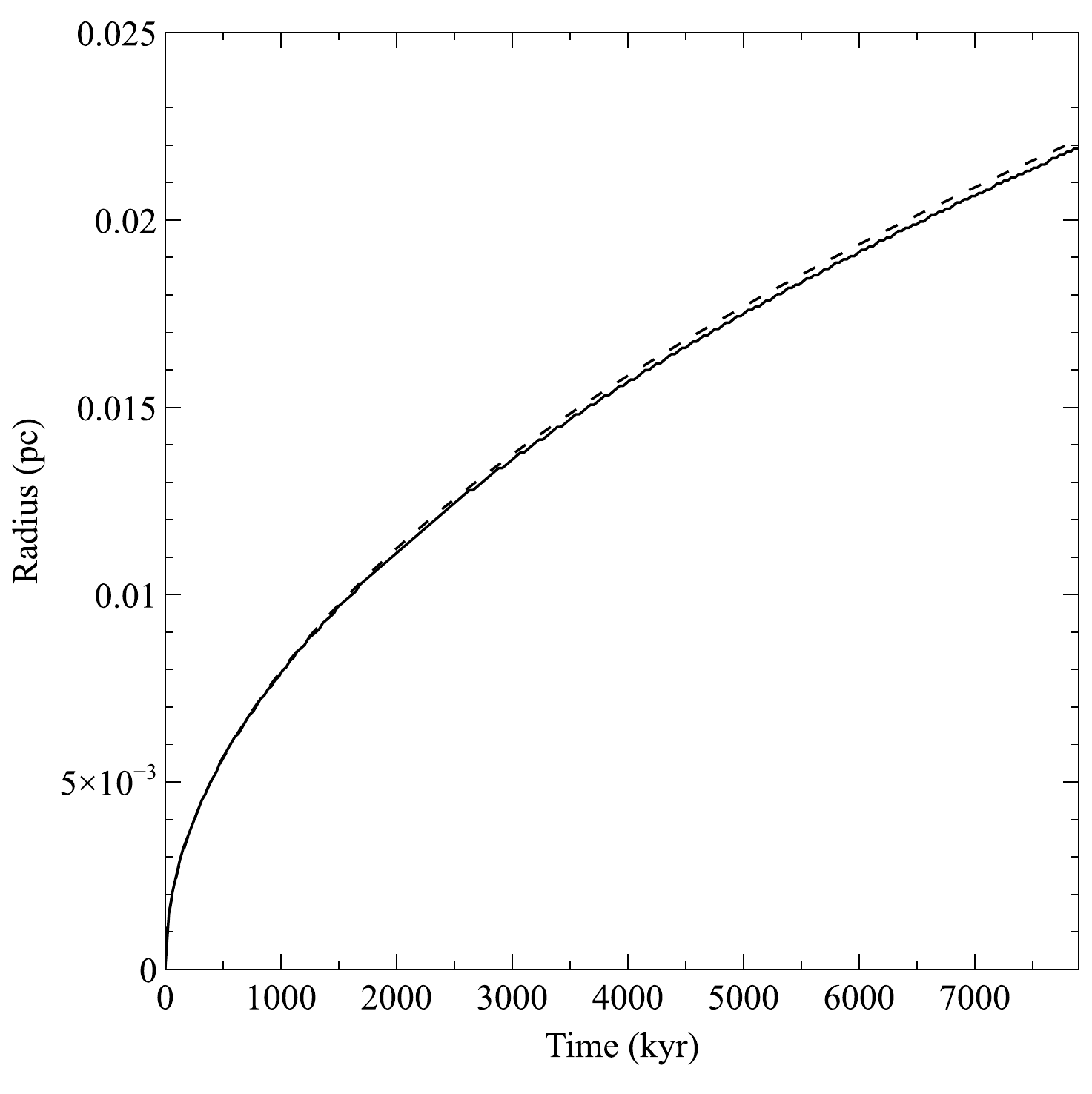}
\caption{The evolution of the shell radius with time for the radiation pressure hydrodynamic test (solid line). The theoretical shell expansion based on the thin-shell approximation is also shown (dashed line).}
\label{radpressure_fig}
\end{figure}

We followed the development of the shell for 8\,kyr, writing out the radial density profile at $10^9$\,s intervals \textbf{(see Figure~\ref{shell_progression_fig})}, and we then determined the position of the shell for each radial profile. This was done by making a parabolic fit to the peak density in the grid and the density at the adjacent radial grid points. We show the results of the dynamical calculation in Figure~\ref{radpressure_fig}. There is good agreement between the model calculation and the theoretical growth of the shell predicted by equation~\ref{shell_eq}. Deviations from theory at late times may be attributed to departures from the thin shell approximation.

\section{Packet splitting}
\label{splitting_sec}

\begin{figure}
\includegraphics[width=80mm]{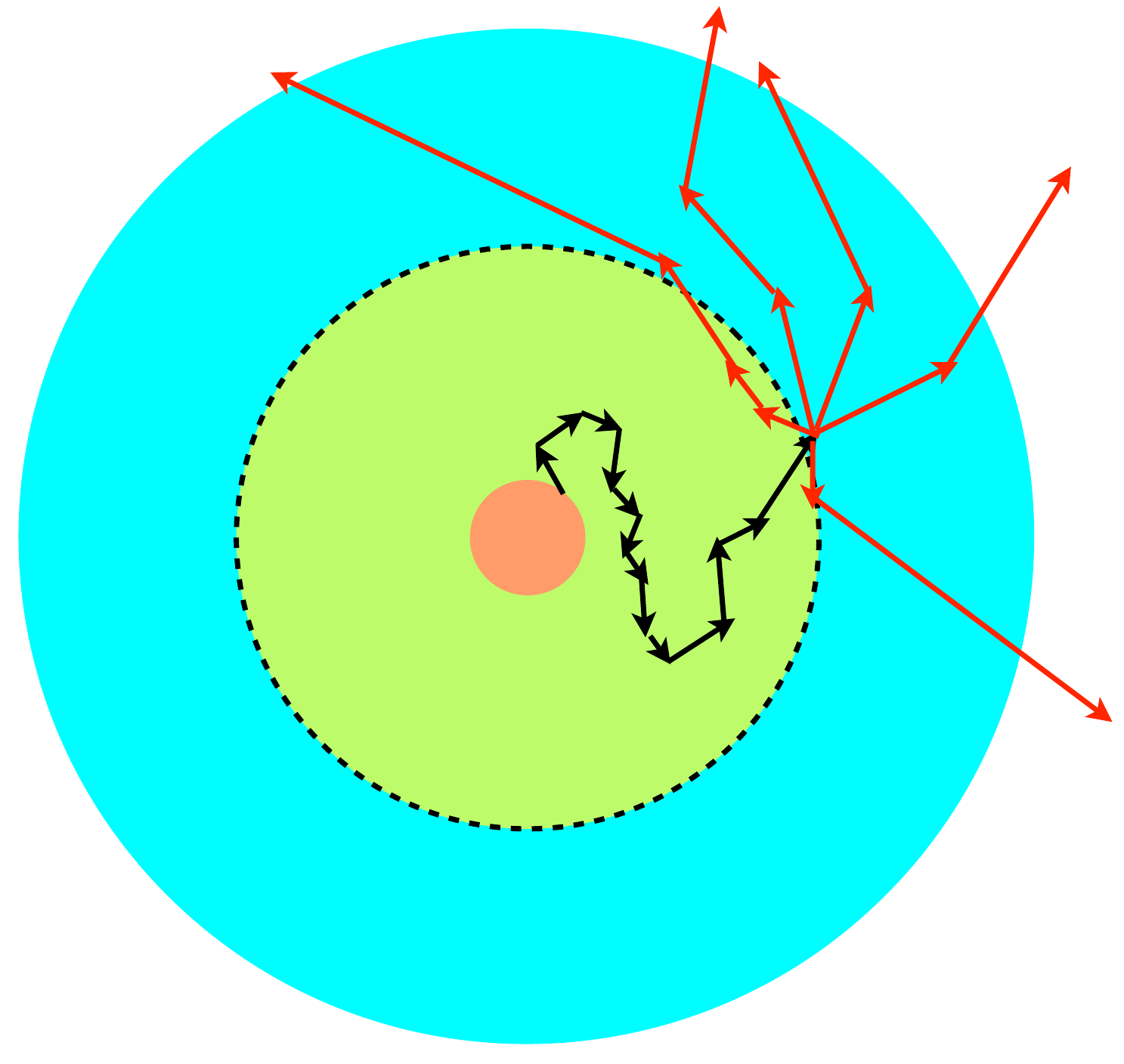}
\caption{A cartoon illustrating the packet splitting method. A high energy packet is released from the protostar and undergoes many absorption and scattering events (black arrows). Eventually the packet passes beyond the region identified as being optically thick (dashed line) and at this point the packet is split into $N_{\rm low}$ low-energy packets (red arrows) which eventually emerge from the computational domain (here $N_{\rm low}=5$). }
\label{splitting_cartoon}
\end{figure}

\textbf{The splitting of particles in the MC method has a long history, dating back to the first neutron transport codes \citep{cashwell_1959}. Splitting, and its inverse, the so-called Russian Roulette method, are variance reduction techniques designed to improve the efficiency with which the MC sampling operates.}  Although the original \cite{lucy_1999} algorithm had equal energy photon packets, this is not a fundamental restriction. In the case of star formation calculations the photon packets may be produced within gas with very high optical depth, and the packets may undergo thousands of scattering events prior to emerging from the computational domain. The MC estimators of the energy density and radiation pressure will be good quality for the optically thick cells where packets spend most of their time, and thus it is inefficient to use equal energy packets. Instead we employ a packet splitting algorithm, in which a lower number of higher energy packets ($N_{\rm high}$) are released from the protostar. These propagate through the optically thick region and then are each split into a number of lower energy packets ($N_{\rm low}$) in the optically thin region (see~\ref{splitting_cartoon}). The total number of packets that emerge from the grid is then $N = N_{\rm high} N_{\rm low}$ The key is to correctly identify the point at which the packet splitting takes place, in order that (i) high energy packets do not propagate into optically thin regions, since they will increase the noise in the MC estimators, (ii) low energy packets undergo a small (but non-zero) number of interactions before leaving the computational domain.

We constructed a one-dimensional test model for the packet splitting
algorithm, comprising a 0.1\, pc radius sphere containing 100\msol\ of
gas and an $r^{-2}$ density profile. The dust opacity was from
\cite{draine_1984} silicates with a dust-to-gas density ratio of 1 per
cent. The dust is heated by a central luminous protostar of $T_{\rm
  eff} = 4000$\,K and radius $R=150$\,${\rm R}_\odot$. We first
calculated the radiative equilibrium without packet splitting using
$N=10^5$ photon packets, and used the final iteration of the radiative
equilibrium calculation as our benchmark.  We subsequently ran the
same model with packet splitting and $N_{\rm high}=1000$ and $N_{\rm
  low}=100$, defining the high optical depth region as that for which
the Planck-mean optical depth to the sphere radius was 20. {\bf Note that in
two or three-dimensional problems the integral of the Planck-mean optical depth is calculated in the positive and negative directions of each coordinate axis, defining a typically ellipsoidal boundary for packet splitting. }

\begin{figure}
\includegraphics[width=80mm]{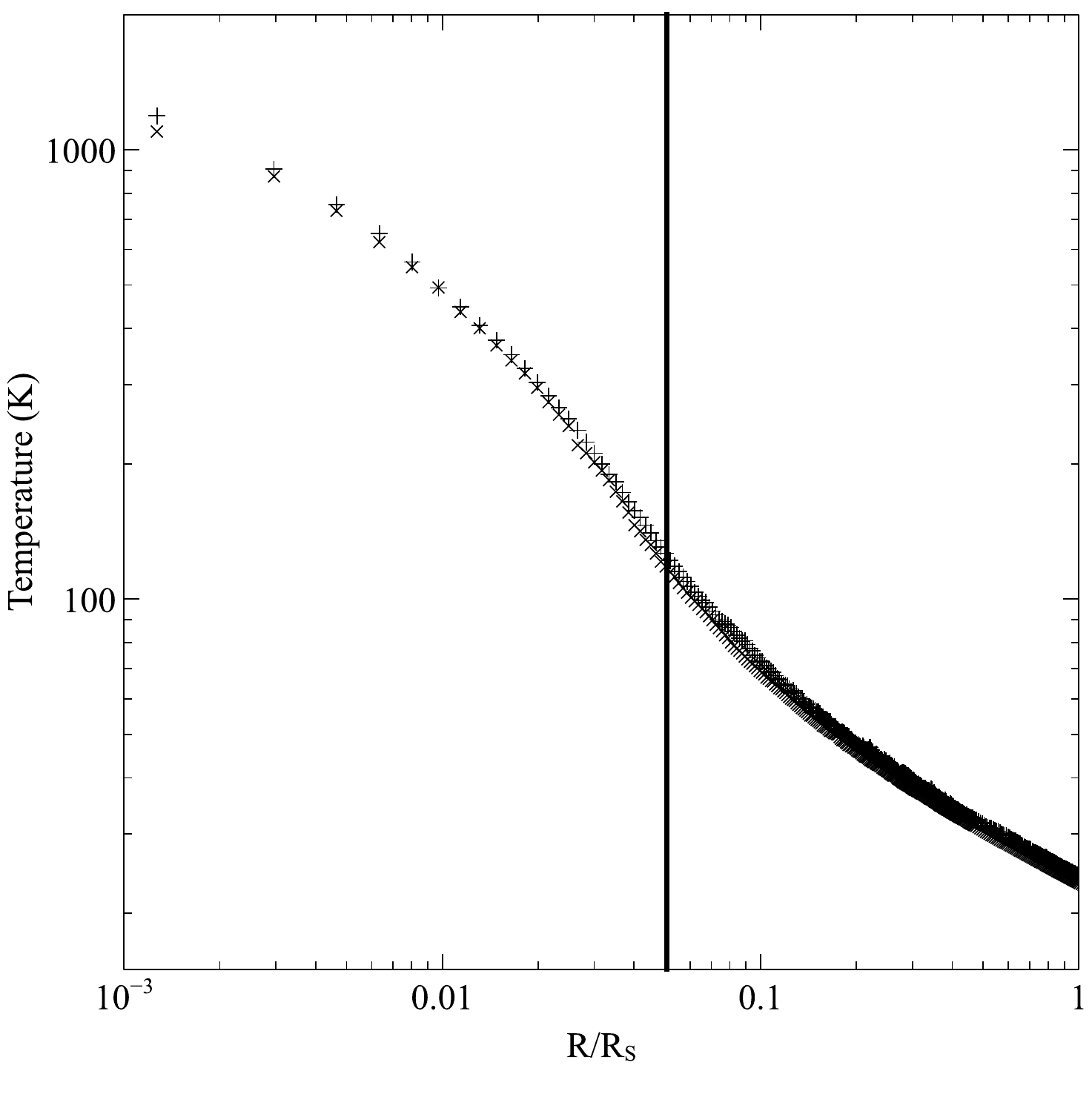}
\caption{The results of the packet splitting test. The temperature of the sphere is plotted as a function of radial distance (in units of the sphere radius $R_s = 0.1$\,pc). The model without packet splitting ($+$ symbol) shows excellent agreement with the temperature found using packet splitting ($\times$ symbol). The vertical line indicates the radius beyond which the high energy packets undergo splitting.}
\label{splitting_fig}
\end{figure}

We plot the temperature profile of the sphere in
Figure~\ref{splitting_fig} for both cases, and find good agreement. In
particular there is a smooth change in the temperature through the
optical depth boundary (the region where the packet splitting occurs).

In the calculation without splitting each photon packet undergoes
$\sim 20\,000$ absorption or scattering events prior to its emergence
from the computational domain.  When packet splitting is enabled each
high energy photon packet still undergoes $\sim 20\,000$ absorptions
or scatterings, but the low energy photon packets only undergo $\sim
200$. This results in a speed-up of the radiation step of a factor of
$\sim 50$.

\section{Parallelisation}
\label{parallel_sec}

The primary drawback in such a detailed treatment of the radiation field is the computational effort involved. Fortunately  the MC method, in which the photon packets are essentially independent events, may be straightforwardly and effectively parallelised. The top level of parallelisation involves domain decomposing the octree that stores the AMR mesh. Each sub-domain belongs to a separate MPI thread, enabling the code to run on distributed memory machines and enabling the use of domains with a larger memory footprint than that available on a single node. We choose a simple domain decomposition, which although not necessarily load balancing, does allow a straightforward implementation in the code. For eight-way decomposition each branch of the octree is stored on an individual MPI process (with an additional thread that performs tasks such as passing photon packets to threads). Similarly 64-way decomposition may be achieved by domain decomposing further down the branches of the octree, and this is the  decomposition that is employed for majority of our runs (although 512-way decomposition is implemented we do not have access to the resources necessary to regularly run the code in this mode). 

The main bottleneck is the communication overhead when passing photon packets between threads, and we optimize this by passing stacks of photon packet data between the MPI threads rather than individual packets in order to reduce latency. \textbf{Thus when a photon packet reaches a boundary between domains (naturally this always corresponds to cell boundary) then the packet position, direction, frequency, and energy of the photon packet is stored on a stack. Once the stack reaches a set number of packets (typically 200), then the stack is passed to the appropriate MPI thread. We note that this algorithm closely resembles the MILAGRO algorithm described by \cite{brunner_2006}. Further optimisation of the stack size is possible, including varying the stack size across the domain boundaries and dynamically altering the stack size as the calculation progresses, but we have yet to implement this.}

\begin{figure}
\includegraphics[width=84mm]{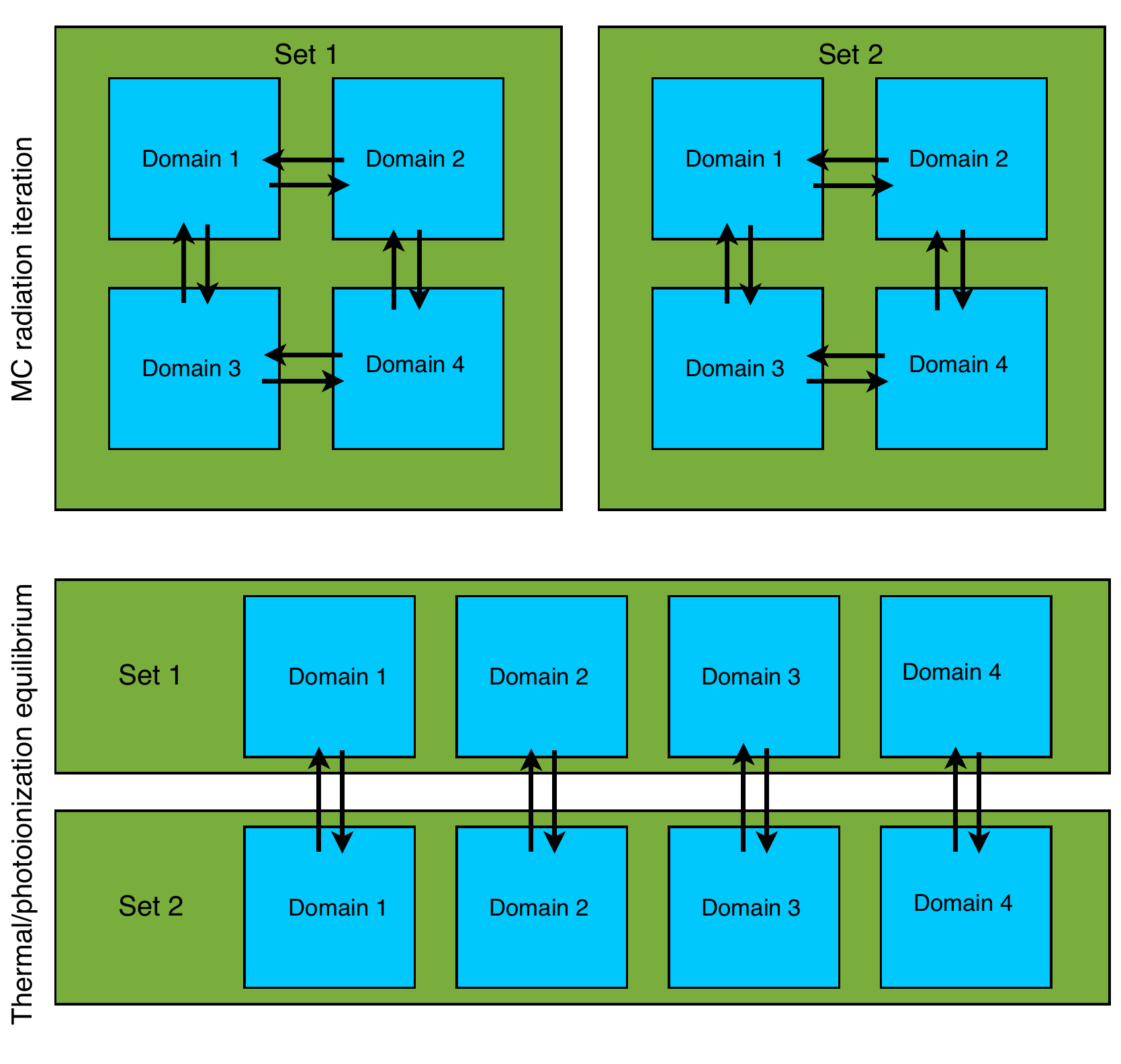}
\caption{A schematic showing the parallelisation of the radiation loop in {\sc torus}. Here the photon packets are split across two sets of the AMR mesh (green squares) and each mesh is divided into four domains (blue squares) , with inter-domain communication of photon packets (indicated by the black arrows). At the end of the iteration the Monte-Carlo estimators are combined for each domain in order to compute revised temperatures and ionization fractions of the gas.}
\label{MPI_chart}
\end{figure}

 A further level of parallelisation is achieved by having many versions of identical computational domains (which we refer to as hydrodynamic sets), over which the photon packet loop is split, with the results derived from the radiation calculation (radiation momenta, cell integrals etc) collated and returned to all the sets at the end of each iteration. The thermal balance and ionization equilibrium calculations (which can be time consuming) are then parallelised over the sets, with each thread corresponding to a particular domain performing the equilibrium over an appropriate fraction of its cells before the results are collated and distributed  (see Figure~\ref{MPI_chart}).

\begin{figure}
\includegraphics[width=80mm]{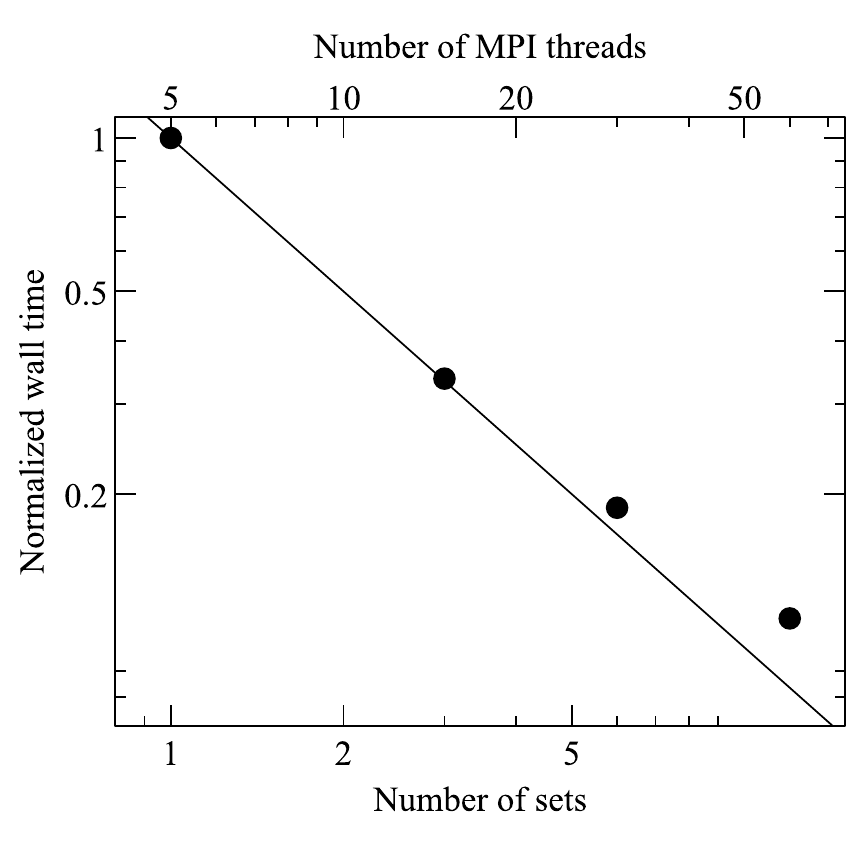}
\caption{Results of the scaling test of the parallelisation. The same 2-D model was run with 1, 3, 6 and 12 hydrodynamics sets, corresponding to 5, 15, 30 and 60 MPI threads. The wall time for one radiation timestep is shown (normalized by the 5-thread calculation). The solid line indicates perfect scaling.}
\label{parallel_fig}
\end{figure}

We performed a scaling test to demonstrate the efficacy of our scheme. We ran the model detailed in section~\ref{splitting_sec}  in 2D. The benchmark used one hydrodynamics set with the quadtree domain-decomposed over 4 threads, giving a total of 5 MPI threads including the control thread. We recorded the wall time for one radiative transfer step, and subsequently ran this same model for 3, 6, and 12 hydrodynamics sets (corresponding to 15, 30 and 60 MPI threads respectively). We found excellent scaling (see Figure~\ref{parallel_fig}) up to 60 threads, with the slight deviation from embarrassing scalability due to the all-to-all MPI communication necessary when the results of the MC estimators are gathered (this is approximately 10 per cent of the wall time for the 60 thread run)‚

\section{Sink particle implementation}

We implemented a Lagrangian sink particle scheme within {\sc torus} in order to follow the formation of stars in our  hydrodynamics code. Originally adopted for smoothed-particle hydrodynamics solvers (\citealt{bate_1995}), sink particles allow one to overcome the restrictively small Courant time associated with high densities as the collapse proceeds, by removing gas from the computation and adding it to point-like particles that interact with the gas via gravity (and possibly radiation), but not via thermal gas pressure.

The fundamental principles for creating a reasonable sink particle implementation are (i) creating sink particles only as a `last resort', (ii) accreting gas in a manner that has minimal impact on the dynamics of the gas immediately around the sink particle, (iii) properly accounting for the gravitational forces between Lagrangian sink particles and the gas on the grid.

The use of sink particles in grid-based hydrodynamics codes has been investigated by \cite{krumholz_2004} and \cite{federrath_2010}.  Broadly speaking the two implementations described deal with items (i) and (iii) in the same manner, but differ in the way that accretion is treated. The Federrath et al. method defines an accretion radius (as a small multiple of the smallest cell size in their AMR mesh) and simply tests that gas in cells within the accretion radius is bound to the sink particle, and then accretes the gas mass (and its associated linear and angular momentum) above a threshold density onto the sink particle. The Krumholz et al. method uses a dynamically varying accretion radius, which ranges from 4-cells to the Bondi-Hoyle radius. The method then ascribes an accretion rate onto the sink based on Bondi-Hoyle-Lyttleton accretion and removes the appropriate mass from each cell within the accretion radius based on a weighting function that falls rapidly with distance from the sink (cells outside the accretion radius have a weighting of zero). The advantage of this method is that an appropriate accretion rate is maintained in situations where the Bondi-Hoyle radius is unresolved, whereas the Federrath method breaks down in this regime. However the Krumholz method relies on a statistical smoothing of the flow in order to determine the appropriate far-field density and velocity to use when calculating the Bondi-Hoyle accretion rate. In contrast the Federrath method relies primarily on the accretion flow being supersonic outside the accretion radius, and therefore the algorithm by which material is removed from the grid has no impact on the upstream dynamics.

In our implementation we use a method akin to Federrath's, since we are always in the regime where the Bondi-Hoyle radius is well resolved. Below we describe the details of our sink particle implementation, and several accretion and dynamics tests of the method.

\subsection{Sink particle creation}

We adopt the same criteria for sink particle creation as \cite{federrath_2010}.
For the cell under consideration we define a control volume that contains all cells within a predefined radius $r_{\rm acc}$.  Before a sink particle is created 
 a number of checks on the hydrodynamical state of the gas in the control volume must be passed, briefly:
\begin{itemize}
\item The central cell of the control volume must have the highest level of AMR refinement.
\item The density of the central cell must exceed a predefined threshold density $\rho_{\rm thresh}$, thus ensuring that ${\bf \nabla} \cdot (\rho {\bf v}) < 0$ for that cell. 
\item Flows in cells along the principle axes must be directed towards the central cell.
\item The gravitation potential of the central cell must be the minimum of all the cells in the control volume.
\item The control volume must be Jeans unstable i.e. $|E_{\rm grav}| > 2E_{\rm th}$.
\item The gas must be in a bound state i.e. $E_{\rm grav} + E_{\rm th} + E_{\rm kin} < 0$, where $E_{\rm kin}$ is the kinetic energy of the gas in the control volume where the speeds are measured relative to the velocity of the centre of mass of the control volume.
\item The control volume must not overlap with the accretion radius of any pre-existing sink particles.
\end{itemize}
If these tests are passed then a sink particle is created at the centre of the control volume, and accretes gas according to the method detailed below.

\subsection{Accretion onto sink particle}

Each sink particle has an associated accretion radius $r_{\rm acc}$ which is defined as 2.5 times the size of the smallest cell in the AMR mesh. The accretion radius defines a spherical volume, and after each hydrodynamics step each the cells in the accretion volume are checked against a predefined density threshold ($\rho_{\rm thresh}$). Cells with a density above $\rho_{\rm thresh}$ undergo a further check to see if the gas in them is  bound to the sink particle, that is
\begin{equation}
E_{\rm grav} + E_{\rm kin} + E_{\rm th} < 0.
\end{equation}
Note the kinetic energy of the gas is measured relative to the velocity of the sink particle. If the gas is both bound and above the density threshold then the mass of the sink particle is increased by $(\rho - \rho_{\rm thresh})V$ where $\rho$ and $V$ are the density and volume of the cell. The linear and angular momentum of the accreted mass is added to the sink particle and subtracted from the cell. 

\begin{figure*}
\includegraphics[width=58mm]{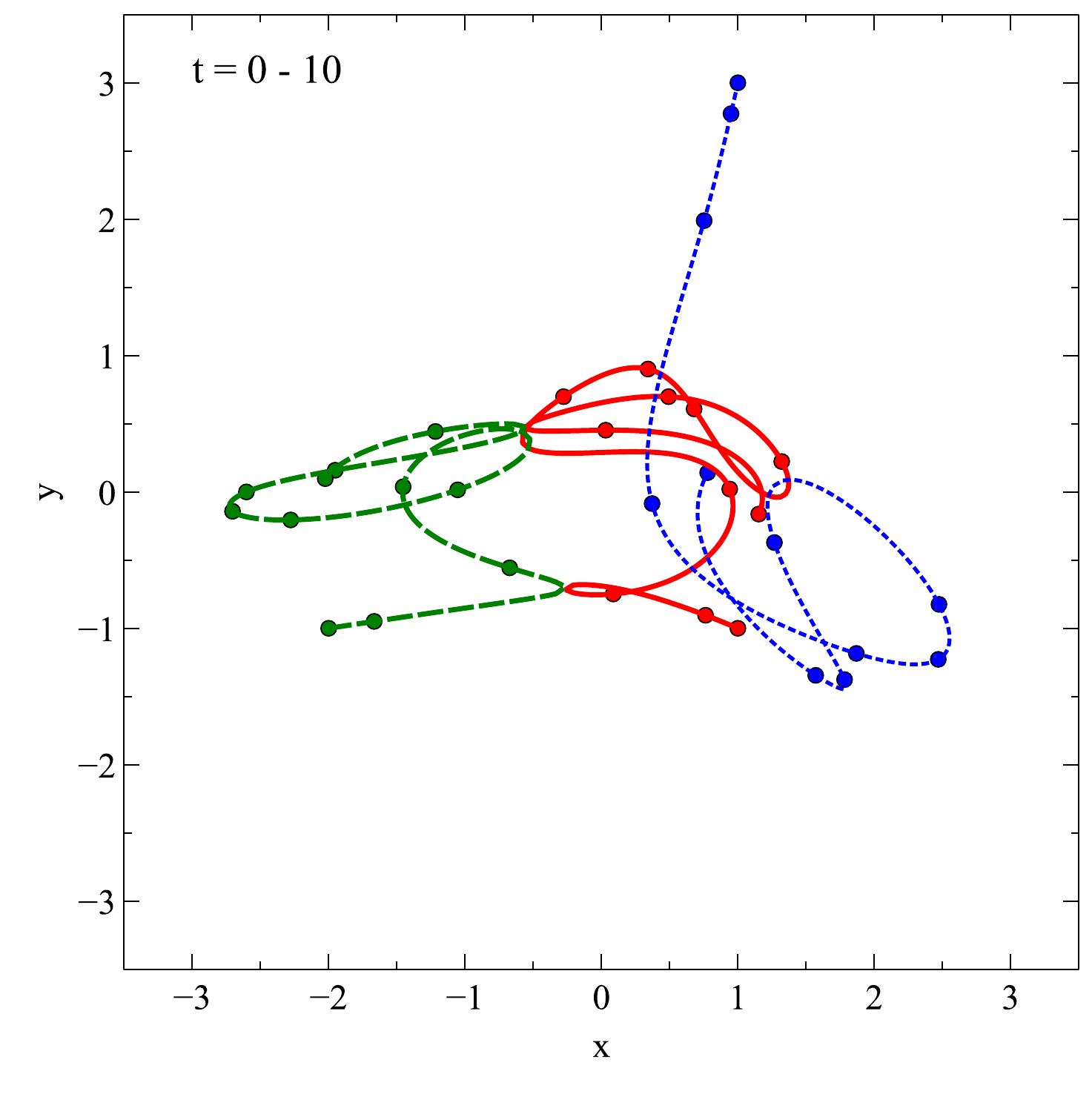}
\includegraphics[width=58mm]{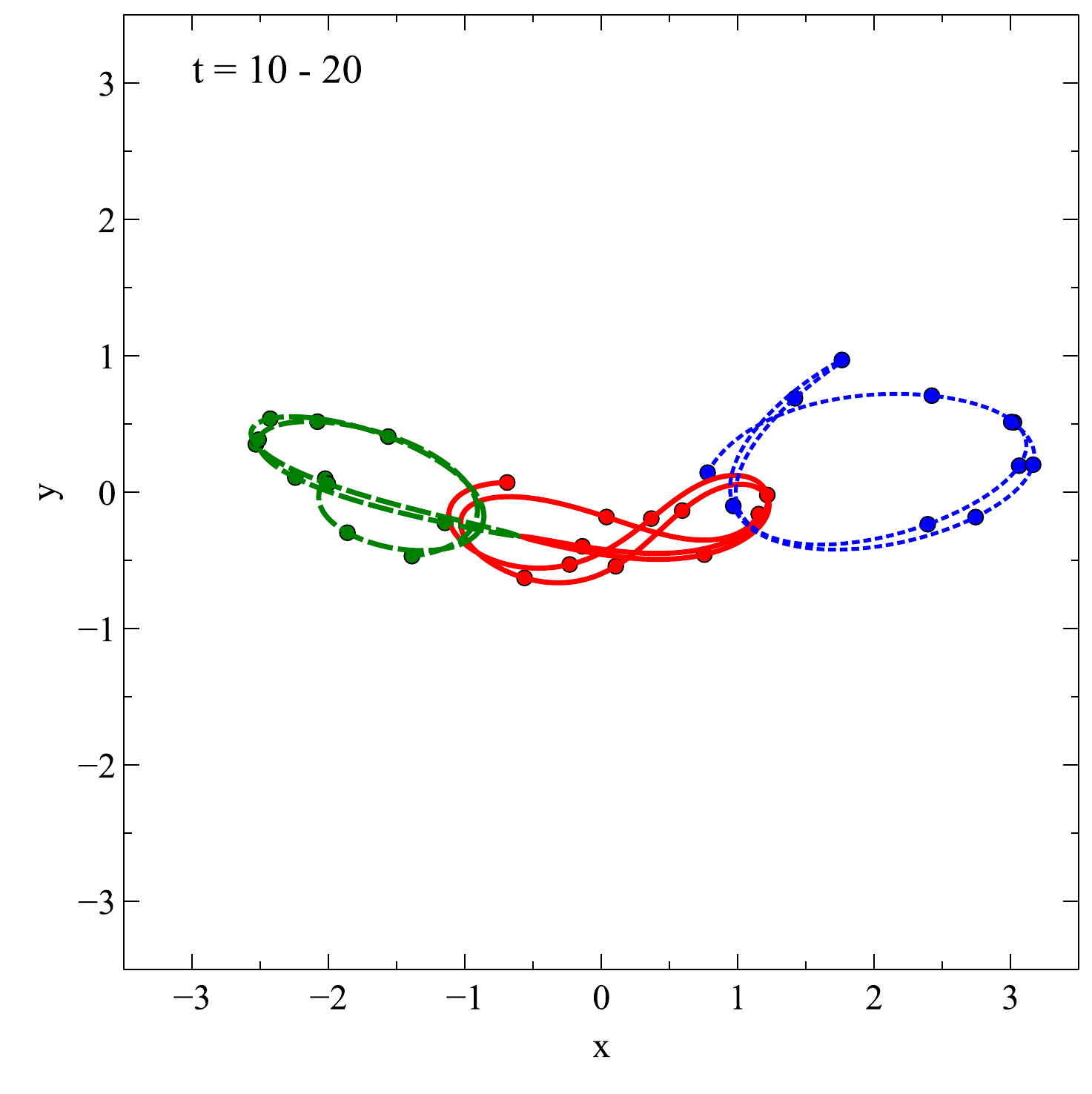}
\includegraphics[width=58mm]{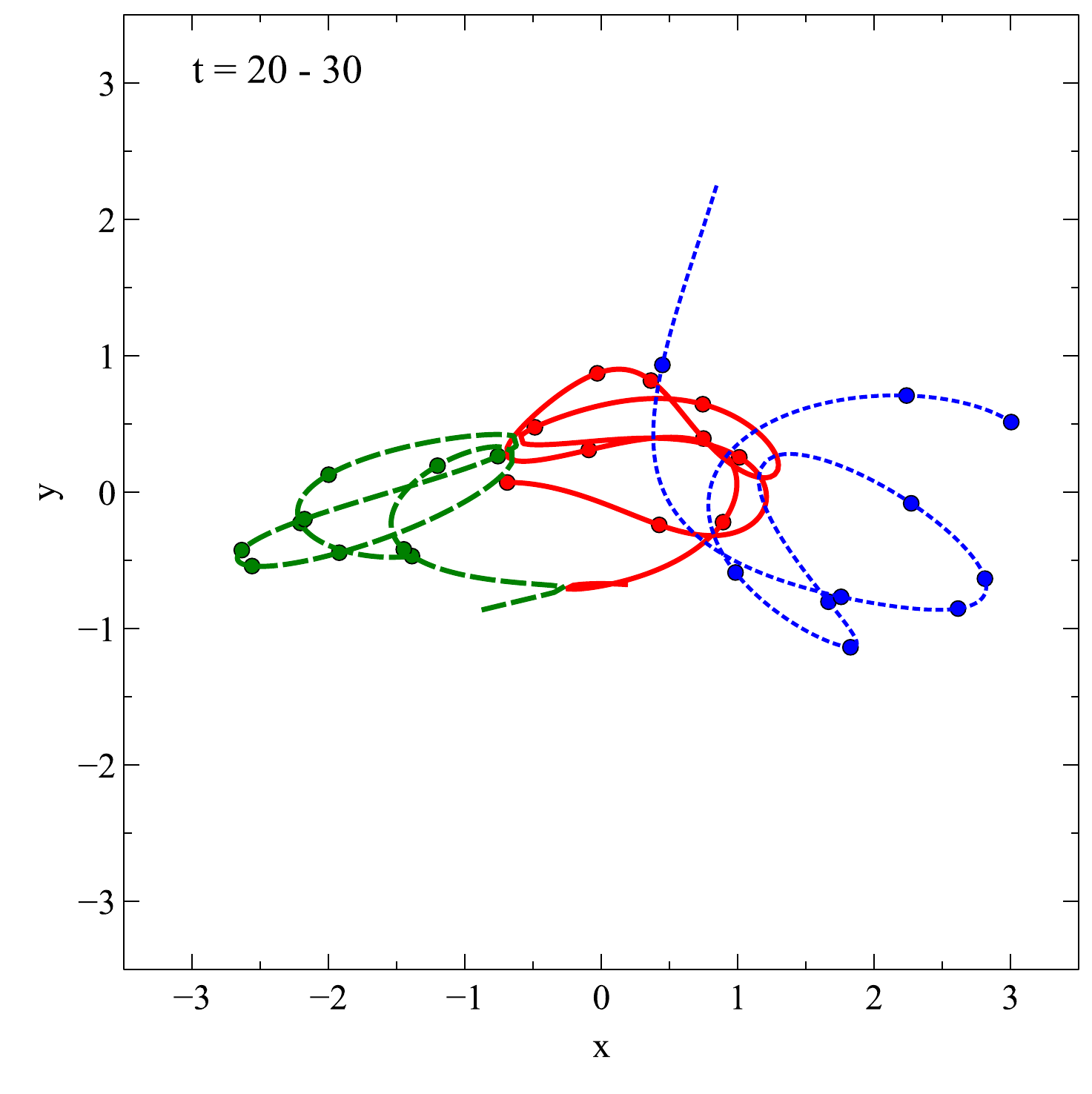}
\caption{The paths of three masses in the Pythagorean three-body benchmark test. The panels shows show the paths of the sink particles at $t=0-10$ (left panel), $t=10-20$ (middle panel), and $t=20-30$ (right panel).  Mass~1 is shown as a green, dashed line, mass~2 as a red, solid line, and mass~3 as a blue, dotted line. Solid circles indicate the position of the particles at unit time intervals.}
\label{threebody_fig}
\end{figure*}

\subsection{Sink particle motion}

The gravitational influence of the gas on a given sink particle is found by summing up the gravitational forces from all the cells in the AMR grid. For all cells excepting that containing the sink particle we use
\begin{equation}
{\bf F}_j = \sum_i G M_j \rho_i V_i  s(r_{ij},h){\bf \hat{r}}
\label{equation_gasforces}
\end{equation} 

where $M_j$ is the mass of the sink particle, and $r_{ij}$ and ${\bf \hat{r}}$ are the distance and direction vector between the cell centre and the sink particle respectively, $h$ is the gravitational softening length, and $s(r_{ij},h)$ is a cubic spine softening function (see equation~A1 of \citealt{price_2007}).

For the cell containing the sink particle we instead perform a sub-grid calculation by splitting the mass of the cell into $8^3$ subcells and sum the force over the subcells in an analogous manner equation~\ref{equation_gasforces}.

The sink-sink interaction is computed using a sum of the gravitational forces over all sinks  (this is computationally tractable since the number of sink particles is small). The equation of motion of the sink particles is integrated over a timestep by using the Bulirsch-Stoer method \citep{press_1993}, once again using the cubic spline softening of \cite{price_2007}. 

\subsection{Gas motion}

The self-gravity of the gas is solved using a multigrid solution to Poisson's equation, and with the gradient of the potential appearing as a source term in hydrodynamics equations.

\section{Sink particle dynamics tests}

Following \cite{hubber_2011} we adopt the three-body Pythagorean test problem of \cite{burrau_1913} to benchmark the sink-sink gravitational interactions and test the integrator. This problem has masses (taking $G=1$) of $m_1 = 3$, $m_2=4$ and $m_3=5$  starting at rest at Cartesian coordinates $(1,3)$, $(-2,-1)$ and ($1,-1)$ respectively. The subsequent motion of the masses  involves a number of close interactions between sinks, and was first numerically integrated by \cite{szebehely_1967}.
The results of the test problem are given in Figure~\ref{threebody_fig} and show excellent agreement  with Figures 2, 3, and 4 of \cite{szebehely_1967}. The total energy of the system over duration of the computation is conserved to better than 1 part in 10$^6$.

\begin{figure}
\includegraphics[width=84mm]{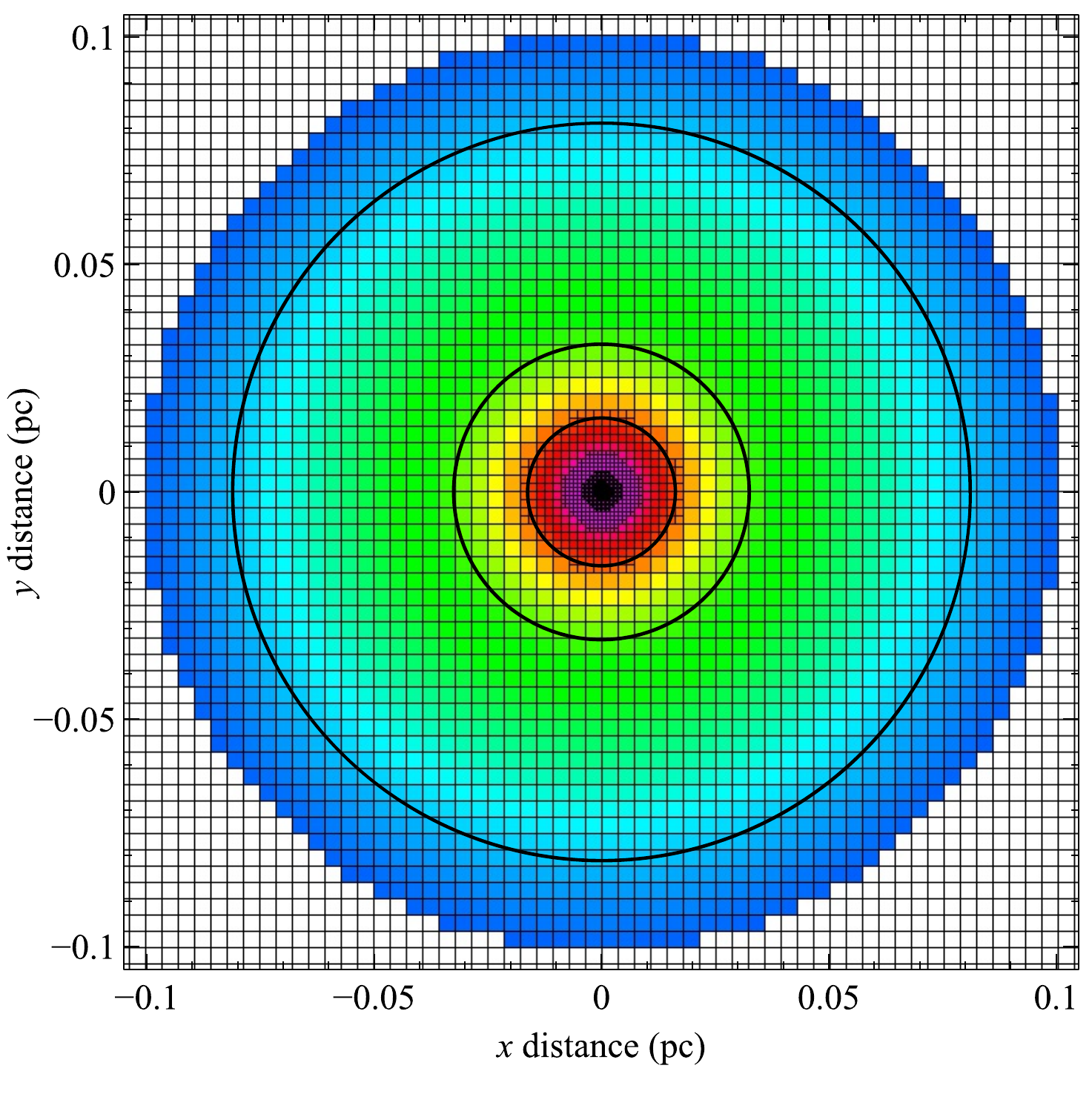}
\begin{center}
log density\\
\includegraphics[width=50mm]{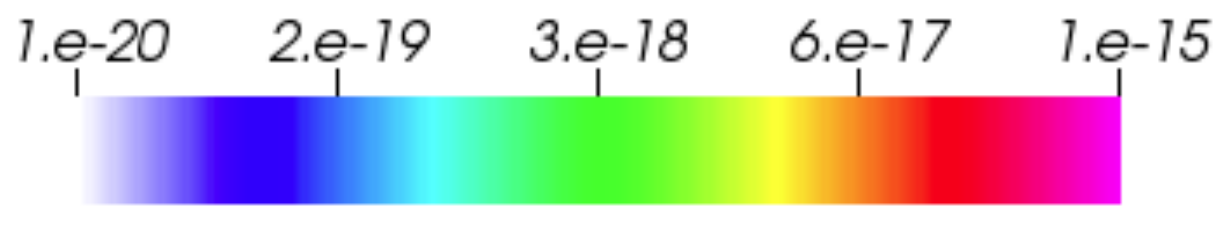}
\end{center}

\caption{The results of the gas-on-sink gravitational force test. The paths of the three test sink particles are shown as bold solid lines, whilst the density of the gas is shown as a logarithmic colour scale. The AMR mesh is shown by thin solid lines.}
\label{nbody_fig}
\end{figure}

The implementation of the gas-on-sink gravitational force calculation was tested by using a power-law ($\rho \propto \rho(r)^{-2}$) density sphere of 100\msol\ and radius 0.1\,pc, on a 3D AMR mesh with minimum cell depth 6 (equivalent to a fixed-grid resolution of $64 \times 64 \times 64$) and maximum cell depth 10 ($1024 \times 1024\times 1024$). The computational domain was divided over 64 MPI threads. Three test sink particles of negligible mass were placed in the grid at radii of $2.5 \times 10^{17}$\,cm, $1 \times 10^{17}$\,cm, and $5 \times 10^{16}$\,cm, at the appropriate Keplerian orbital speed ($\sim 2.074$\,km\,s$^{-1}$). The hydrodynamics of the gas was neglected for this test, and the sink particles were allowed to move through the stationary gas under gravitational forces only. The benchmark test was run for 130 orbits of the inner-most particle (corresponding to $\sim 26$ orbits of the outermost particle). The orbits of the particles are overlaid on the density distribution and adaptive mesh in Figure~\ref{nbody_fig}, indicating that the integrator, domain decomposition, and gas-particle forces are operating satisfactorily.

\section{Accretion tests}

The principle tests of our sink particle implementation are those that ensure that the accretion of gas occurs at the correct rate. We therefore conducted three benchmark tests of increasing complexity.

\subsection{Bondi accretion}

A point mass $M$ accretes spherically from a gas cloud, where far from $M$ the cloud is stationary and has density $\rho_\infty$.  \cite{bondi_1952} showed that the maximum accretion rate is given by
\begin{equation}
\dot{M} = 4 \pi \lambda \frac{G^2M^2}{c_s^3}\rho_\infty
\end{equation}
where $c_s$ is the sound speed and $\lambda=1.12$ for isothermal gas. The corresponding Bondi radius is given by
\begin{equation}
R_B = \frac{2GM}{c_s^2}.
\end{equation}

\begin{figure}
\includegraphics[width=80mm]{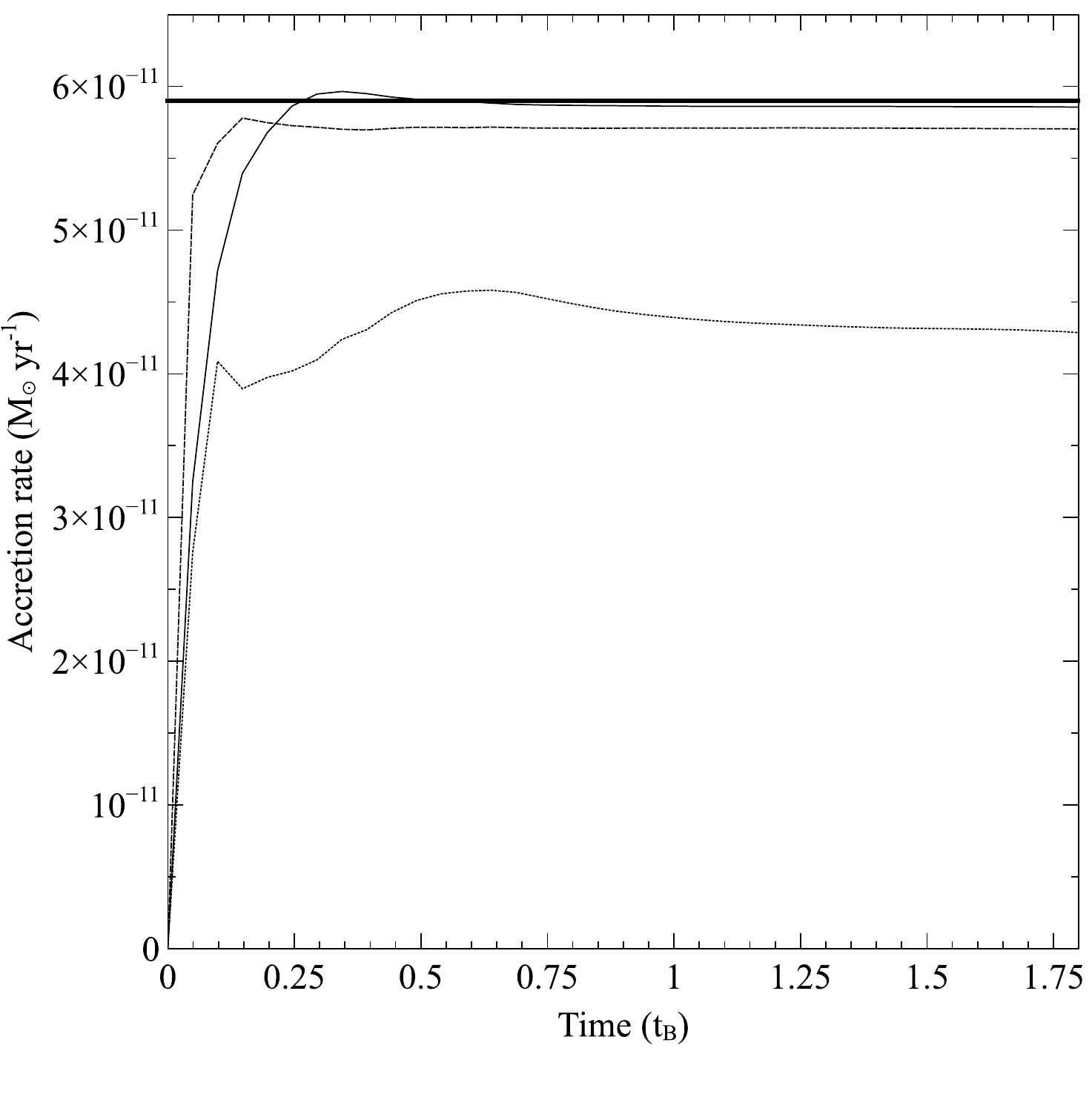}
\caption{Results of the Bondi accretion test. The accretion rate is plotted against the Bondi timescale ($t_B = R_B / c_s$) for the $R_B/\Delta x$ = 5, 10, and 20 models (dotted, dashed and solid lines repsectively). The thick solid line corresponds to the expected theoretical rate of $5.9 \times 10^{-11}$ \mdotrate.}
\label{bondi_fig}
\end{figure}

For this test we adopted $M=1$\,\msol\ and $\rho_\infty = 10^{-25}$\,g\,cc$^{-1}$, giving an expected accretion rate of $5.9 \times 10^{-11}$ \mdotrate and a Bondi radius of $3.75 \times 10^{17}$\,cm. We ran three two-dimensional models with resolutions of $R_B/\Delta x$ of 5, 10, and 20. Rapid convergence in the accretion rate with resolution is observed (see Figure~\ref{bondi_fig}). The highest resolution model has an accretion rate of $5.7 \times 10^{-11}$ \mdotrate, i.e. within 2\% of the theoretical rate. 

\subsection{Collapse of a singular isothermal sphere}

\cite{shu_1977} showed that an isothermal sphere with $\rho(r) \propto 1/r^2$ collapses in such a way that there is a constant mass flux through spherical shells. The added level of complexity over the Bondi test described above is that the self-gravity of the gas is significant.

We adopt the same test as  \cite{federrath_2010}, with a sphere of radius $R = 5 \times 10^{16}$\,cm with $\rho  = 3.82 \times 10^{-18}$\,g\,cm$^{-3}$ with contains 3.02\msol\ of gas. The sound speed of the gas was taken to be 0.166\,km\,s$^{-1}$. The expected accretion rate of this model is $1.5 \times 10^{-4}$\mdotrate. We adopted an adaptive, two-dimensional cylindrical mesh for four levels of refinement, with the smallest cells of  $R/\Delta x = 300$

\begin{figure}
\includegraphics[width=84mm]{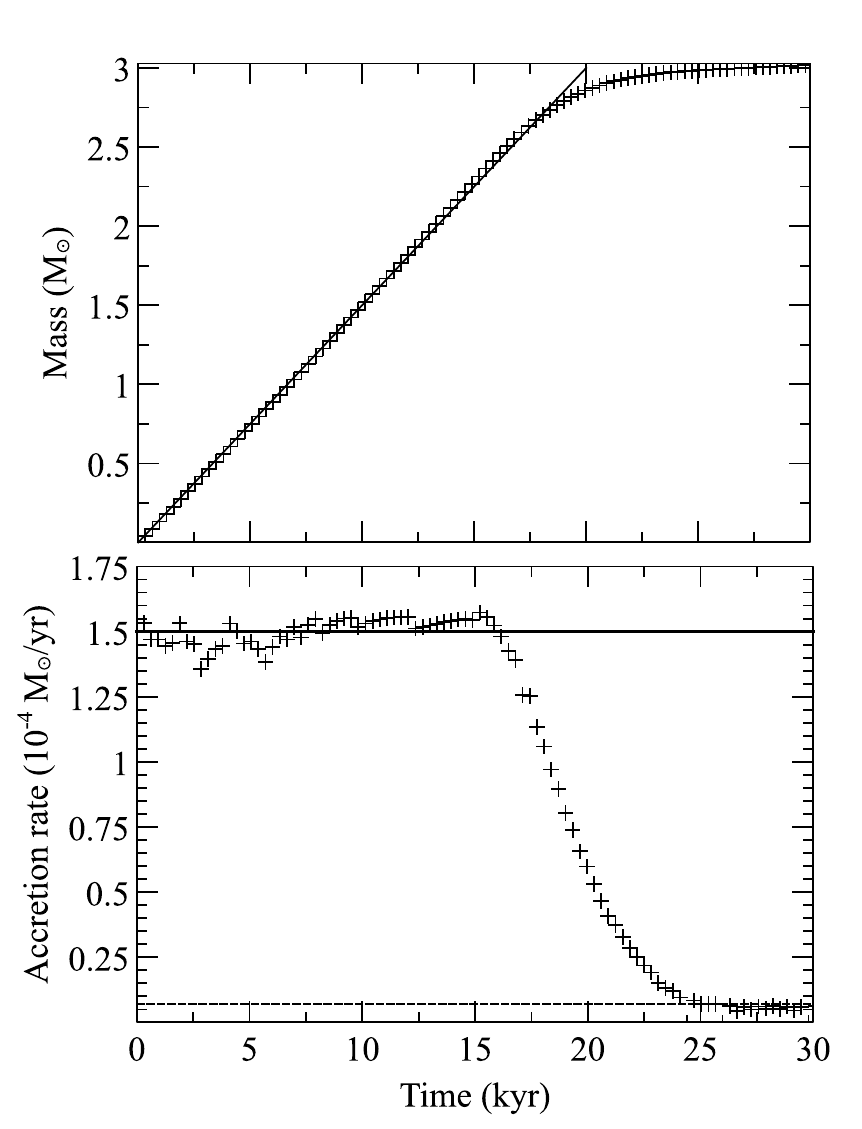}
\caption{Results of the Shu accretion test. The lower panel shows the accretion rate as a function of time (crosses), with the expected theoretical accretion rates predicted by the Shu model (solid line) and Bondi accretion at the floor density (dashed line). The upper panel shows the growth of the central object as a function of time (crosses) along with the expected trend assuming a constant Shu accretion rate (solid line).}
\label{shu_fig}
\end{figure}

The model demonstrated excellent agreement with the Shu prediction, starting with an accretion rate of $\sim 1.5 \times 10^{-4}$\mdotrate, and only declining when 90\% of the original mass had been accreted (see Figure~\ref{shu_fig}). At the end of the run the local density approached the global floor density for the simulation ($10^{-21}$\,g\,cc$^{-1}$) and the accretion rate approaches the Bondi rate for that density.

\subsection{Bondi-Hoyle accretion}

We constructed a 2D test case in which a 1\,\msol\ sink is placed at the origin in initially uniform density gas ($10^{-25}$\,g\,cc$^{-1}$) with molecular weight of 2.33 and a temperature of 10\,K. The gas initially had a constant velocity with a mach number of ${\cal M}=3$ parallel to the $z$-axis, and an inflow condition at the upstream boundary and and outflow condition at the downstream boundary. These are the same initial conditions as the Bondi-Hoyle test case in \cite{krumholz_2004}, although they ran their simulation in 3D.

\begin{figure}
\includegraphics[width=80mm]{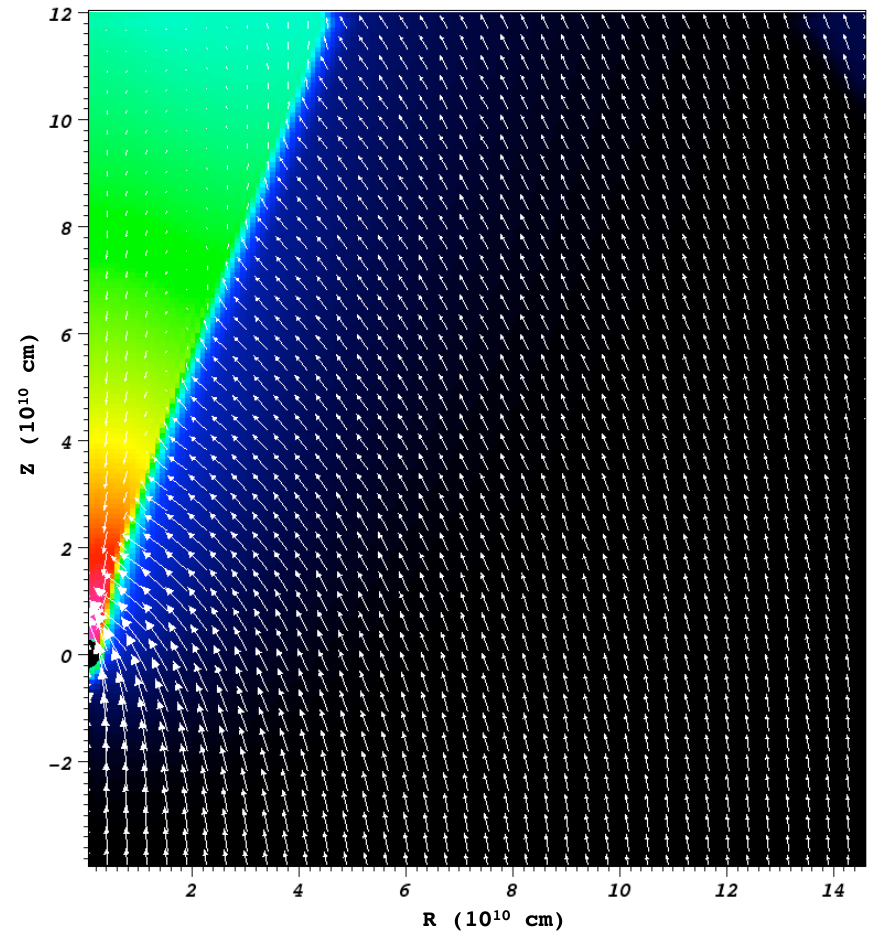}
\caption{Bondi-Hoyle accretion test. The figure shows logarithmically scaled density between $10^{-23}$ g\,cc$^{-1}$ (pink) and $10^{-25}$ g\,cc$^{-1}$ (black). The white arrows show velocity vectors, with the longest arrows corresponding to speeds of 3\,km\,s$^{-1}$. The black semicircular region at the origin signifies  the extent of the accretion region.}
\label{bondihoyle_dens_fig}
\end{figure}

\begin{figure}
\includegraphics[width=80mm]{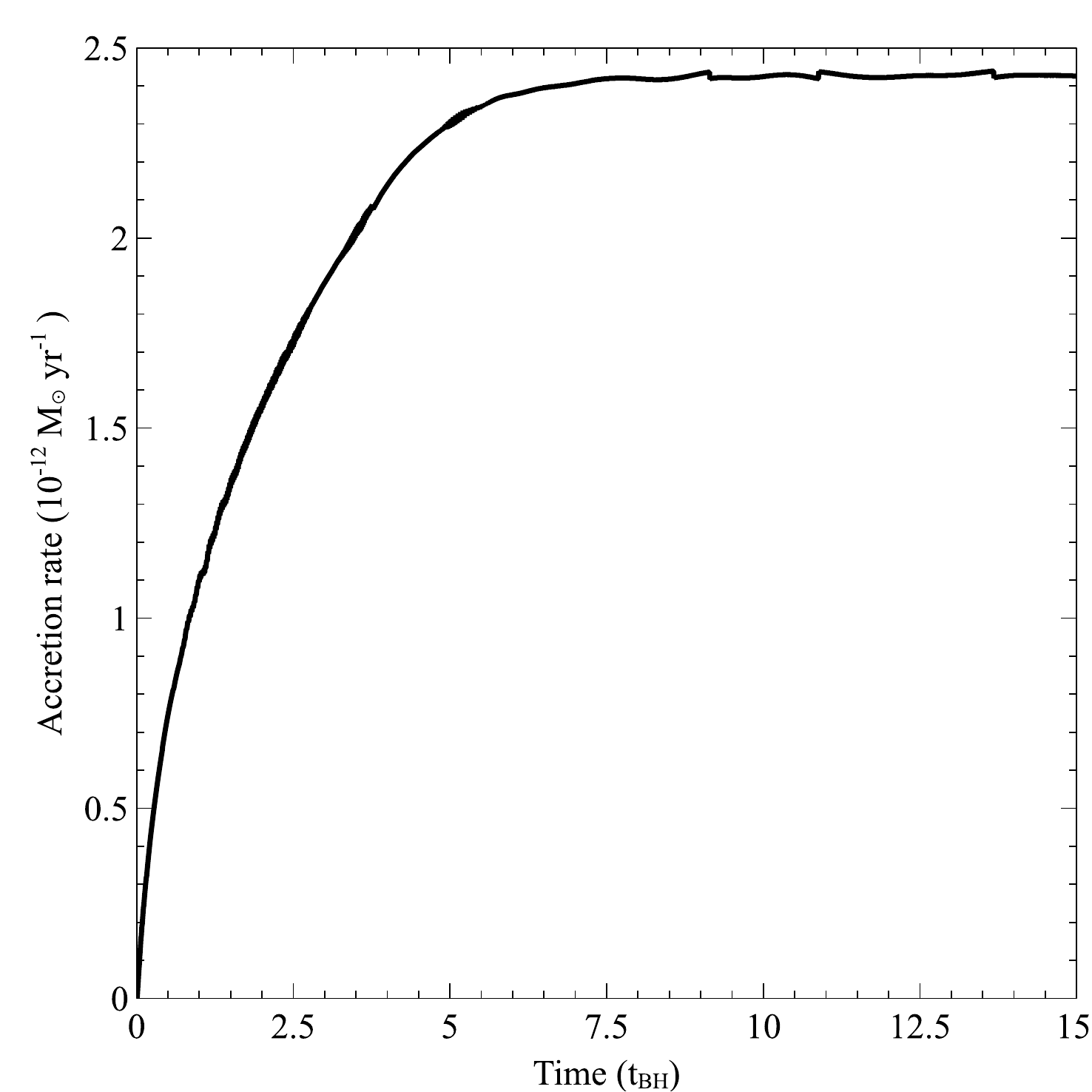}
\caption{Bondi-Hoyle accretion as a function of time. Time is given in units of the Bondi-Hoyle timescale, while the accretion rate is given in units of $10^{-12}$ \mdotrate. }
\label{bondihoyle_acc_fig}
\end{figure}

The model extent was  0.78\,pc, and three levels of refinement were used with the smallest cells corresponding to $2.3 \times 10^{14}$\,cm. We ran the model until an approximately steady-state of the accretion rate was achieved (Figure~\ref{bondihoyle_dens_fig}). The theoretical Bondi-Hoyle accretion rate for these conditions is $1.7 \times 10^{-12}$\mdotrate, but the \cite{krumholz_2004} model, and those of \cite{ruffert_1996}, found accretion rates of close to $2 \times 10^{-12}$\mdotrate, with considerable temporal variation in the accretion rate on the Bondi-Hoyle timescale. Our simulation reached a steady-state accretion rate of $2.4 \times 10^{-12}$\mdotrate (Figure~\ref{bondihoyle_acc_fig}), which is comparable to the peak accretion rate seen in the \cite{krumholz_2004} simulations. The level of variability is substantially lower in our simulation, presumably because the instabilities that build up and modify the dynamics near the sink in the 3D simulations are absent in our 2D models. 

\section{Conclusions}

We have presented a new method for including radiation pressure in RHD simulations that incorporates a level of microphysical detail that is significantly greater than that of flux-limited diffusion or hybrid techniques. We have shown that the new method works in both the pure absorption and pure scattering regimes, and properly treats anisotropic scattering processes. The method comprises a simple addition to the MC estimators required for radiative and photoionisation equilibrium calculations, and does not therefore represent a substantial computational overhead to the MC RHD methods described by \cite{haworth_2012a}. 

However the MC method as a whole {\em is} significantly more computationally demanding than the FLD and hybrid methods, and we have therefore developed two new methods to ameliorate this. The first is the packet splitting method detailed in section~\ref{splitting_sec}, in which we extend Lucy's original algorithm to incorporate photon packets with varying energy. The second is to distribute the MC photon packet loop over many instances of the computational domain, allowing an excellent scalability (see section~\ref{parallel_sec}).

The final element needed to compute massive star formation models is a description of the protostar itself, and to do this we have included a sink particle algorithm. By interpolating on protostellar evolutionary model grids (e.g. \citealt{hosokawa_2009}) as a function of mass and accretion rate will will be able to determine temperatures and luminosities for our protostar, and assign it a spectral energy distribution from appropriate atmosphere models. The sink particles  then become the origin of photon packets for the RHD calculation. 

The next stage is to compute massive star formation models that incorporate radiation pressure and ionisation feedback. Initially these models will be two dimensional, but we also  conduct three-dimensional calculations in order to simulate binary star formation. Of course the algorithms detailed here have wider applicability, and the correct treatment of feedback from ionisation, radiation pressure and winds is important for cluster-scale calculations, in particular with reference to gas dispersal from clusters (e.g. \citealt{rogers_2013}).

\section*{Acknowledgments}

The calculations for this paper were performed on the University of Exeter Supercomputer, a DiRAC Facility jointly funded by STFC, the Large Facilities Capital fund of BIS, and the University of Exeter, and on the Complexity DiRAC Facility jointly funded by STFC and the Large Facilities Capital Fund of BIS. TJH acknowledges funding from Exeter's STFC Consolidated Grant (ST/J001627/1). \textbf{I am grateful to Jon Bjorkman for useful advice on forward scattering given over cocktails provided by Kenny Wood.} Dave Acreman, Tom Haworth,  Matthew Bate, and Ant Whitworth are thanked for useful discussions.

\bsp

\label{lastpage}
\bibliographystyle{mn2e}
\bibliography{torus}

\begin{thebibliography}{}

\bibitem[\protect\citeauthoryear{{Acreman}, {Harries} \& {Rundle}}{{Acreman}
  et~al.}{2010}]{acreman_2010}
{Acreman} D.~M.,  {Harries} T.~J.,    {Rundle} D.~A.,  2010, MNRAS, 403, 1143

\bibitem[\protect\citeauthoryear{{Bate}, {Bonnell} \& {Price}}{{Bate}
  et~al.}{1995}]{bate_1995}
{Bate} M.~R.,  {Bonnell} I.~A.,    {Price} N.~M.,  1995, MNRAS, 277, 362

\bibitem[\protect\citeauthoryear{{Bjorkman} \& {Wood}}{{Bjorkman} \&
  {Wood}}{2001}]{bjorkman_2001}
{Bjorkman} J.~E.,  {Wood} K.,  2001, ApJ, 554, 615

\bibitem[\protect\citeauthoryear{{Bondi}}{{Bondi}}{1952}]{bondi_1952}
{Bondi} H.,  1952, MNRAS, 112, 195

\bibitem[\protect\citeauthoryear{Brunner, Urbatsch, Evans \& Gentile}{Brunner
  et~al.}{2006}]{brunner_2006}
Brunner T.~A.,  Urbatsch T.~J.,  Evans T.~M.,    Gentile N.~A.,  2006, J.
  Comput. Phys., 212, 527

\bibitem[\protect\citeauthoryear{{Burrau}}{{Burrau}}{1913}]{burrau_1913}
{Burrau} C.,  1913, Astronomische Nachrichten, 195, 113

\bibitem[\protect\citeauthoryear{Cashwell \& Everett}{Cashwell \&
  Everett}{1959}]{cashwell_1959}
Cashwell E.~D.,  Everett C.~J.,  1959, A practical manual on the Monte Carlo
  method for random walk problems, 1st edn.
Permagon Press, London

\bibitem[\protect\citeauthoryear{{Draine} \& {Lee}}{{Draine} \&
  {Lee}}{1984}]{draine_1984}
{Draine} B.~T.,  {Lee} H.~M.,  1984, ApJ, 285, 89

\bibitem[\protect\citeauthoryear{{Federrath}, {Banerjee}, {Clark} \&
  {Klessen}}{{Federrath} et~al.}{2010}]{federrath_2010}
{Federrath} C.,  {Banerjee} R.,  {Clark} P.~C.,    {Klessen} R.~S.,  2010, ApJ,
  713, 269

\bibitem[\protect\citeauthoryear{{Ferland}, {Porter}, {van Hoof}, {Williams},
  {Abel}, {Lykins}, {Shaw}, {Henney} \& {Stancil}}{{Ferland}
  et~al.}{2013}]{ferland_2013}
{Ferland} G.~J.,  {Porter} R.~L.,  {van Hoof} P.~A.~M.,  {Williams} R.~J.~R.,
  {Abel} N.~P.,  {Lykins} M.~L.,  {Shaw} G.,  {Henney} W.~J.,    {Stancil}
  P.~C.,  2013, Revista Mexicana de Astronomía y Astrofísica, 49, 137

\bibitem[\protect\citeauthoryear{{Harries}}{{Harries}}{2011}]{harries_2011}
{Harries} T.~J.,  2011, MNRAS, 416, 1500

\bibitem[\protect\citeauthoryear{{Haworth} \& {Harries}}{{Haworth} \&
  {Harries}}{2012}]{haworth_2012a}
{Haworth} T.~J.,  {Harries} T.~J.,  2012, MNRAS, 420, 562

\bibitem[\protect\citeauthoryear{{Haworth}, {Harries} \& {Acreman}}{{Haworth}
  et~al.}{2012}]{haworth_2012b}
{Haworth} T.~J.,  {Harries} T.~J.,    {Acreman} D.~M.,  2012, MNRAS, 426, 203

\bibitem[\protect\citeauthoryear{{Haworth}, {Harries}, {Acreman} \&
  {Rundle}}{{Haworth} et~al.}{2013}]{haworth_2013}
{Haworth} T.~J.,  {Harries} T.~J.,  {Acreman} D.~M.,    {Rundle} D.~A.,  2013,
  MNRAS, 431, 3470

\bibitem[\protect\citeauthoryear{{Hosokawa} \& {Omukai}}{{Hosokawa} \&
  {Omukai}}{2009}]{hosokawa_2009}
{Hosokawa} T.,  {Omukai} K.,  2009, ApJ, 691, 823

\bibitem[\protect\citeauthoryear{{Hubber}, {Batty}, {McLeod} \&
  {Whitworth}}{{Hubber} et~al.}{2011}]{hubber_2011}
{Hubber} D.~A.,  {Batty} C.~P.,  {McLeod} A.,    {Whitworth} A.~P.,  2011,
  A\&A, 529, A27

\bibitem[\protect\citeauthoryear{{Kennicutt}
  Jr.}{{Kennicutt}}{1998}]{kennicutt_1998}
{Kennicutt} Jr. R.~C.,  1998, {ARAA}, 36, 189

\bibitem[\protect\citeauthoryear{{Krumholz}, {Klein}, {McKee}, {Offner} \&
  {Cunningham}}{{Krumholz} et~al.}{2009}]{krumholz_2009}
{Krumholz} M.~R.,  {Klein} R.~I.,  {McKee} C.~F.,  {Offner} S.~S.~R.,
  {Cunningham} A.~J.,  2009, Science, 323, 754

\bibitem[\protect\citeauthoryear{{Krumholz}, {McKee} \& {Klein}}{{Krumholz}
  et~al.}{2004}]{krumholz_2004}
{Krumholz} M.~R.,  {McKee} C.~F.,    {Klein} R.~I.,  2004, ApJ, 611, 399

\bibitem[\protect\citeauthoryear{{Kuiper}, {Klahr}, {Beuther} \&
  {Henning}}{{Kuiper} et~al.}{2010}]{kuiper_2010b}
{Kuiper} R.,  {Klahr} H.,  {Beuther} H.,    {Henning} T.,  2010, ApJ, 722, 1556

\bibitem[\protect\citeauthoryear{{Kuiper}, {Klahr}, {Beuther} \&
  {Henning}}{{Kuiper} et~al.}{2012}]{kuiper_2012}
{Kuiper} R.,  {Klahr} H.,  {Beuther} H.,    {Henning} T.,  2012, A\&A, 537,
  A122

\bibitem[\protect\citeauthoryear{{Kuiper}, {Klahr}, {Dullemond}, {Kley} \&
  {Henning}}{{Kuiper} et~al.}{2010}]{kuiper_2010a}
{Kuiper} R.,  {Klahr} H.,  {Dullemond} C.,  {Kley} W.,    {Henning} T.,  2010,
  A\&A, 511, A81

\bibitem[\protect\citeauthoryear{{Kuiper} \& {Klessen}}{{Kuiper} \&
  {Klessen}}{2013}]{kuiper_2013}
{Kuiper} R.,  {Klessen} R.~S.,  2013, A\&A, 555, A7

\bibitem[\protect\citeauthoryear{{Lucy}}{{Lucy}}{1999}]{lucy_1999}
{Lucy} L.~B.,  1999, A\&A, 344, 282

\bibitem[\protect\citeauthoryear{{Nayakshin}, {Cha} \& {Hobbs}}{{Nayakshin}
  et~al.}{2009}]{nayakshin_2009}
{Nayakshin} S.,  {Cha} S.-H.,    {Hobbs} A.,  2009, ArXiv e-prints

\bibitem[\protect\citeauthoryear{{Noebauer}, {Sim}, {Kromer}, {R{\"o}pke} \&
  {Hillebrandt}}{{Noebauer} et~al.}{2012}]{noebauer_2012}
{Noebauer} U.~M.,  {Sim} S.~A.,  {Kromer} M.,  {R{\"o}pke} F.~K.,
  {Hillebrandt} W.,  2012, MNRAS, 425, 1430

\bibitem[\protect\citeauthoryear{Press, Teukolsky, Vetterling \&
  Flannery}{Press et~al.}{1993}]{press_1993}
Press W.~H.,  Teukolsky S.~A.,  Vetterling W.~T.,    Flannery B.~P.,  1993,
  Numerical Recipes in FORTRAN; The Art of Scientific Computing, 2nd edn.
Cambridge University Press, New York, NY, USA

\bibitem[\protect\citeauthoryear{{Price} \& {Monaghan}}{{Price} \&
  {Monaghan}}{2007}]{price_2007}
{Price} D.~J.,  {Monaghan} J.~J.,  2007, MNRAS, 374, 1347

\bibitem[\protect\citeauthoryear{{Rogers} \& {Pittard}}{{Rogers} \&
  {Pittard}}{2013}]{rogers_2013}
{Rogers} H.,  {Pittard} J.~M.,  2013, MNRAS, 431, 1337

\bibitem[\protect\citeauthoryear{{Roth} \& {Kasen}}{{Roth} \&
  {Kasen}}{2014}]{roth_2014}
{Roth} N.,  {Kasen} D.,  2014, ArXiv e-prints

\bibitem[\protect\citeauthoryear{{Ruffert}}{{Ruffert}}{1996}]{ruffert_1996}
{Ruffert} M.,  1996, A\&A, 311, 817

\bibitem[\protect\citeauthoryear{{Rundle}, {Harries}, {Acreman} \&
  {Bate}}{{Rundle} et~al.}{2010}]{rundle_2010}
{Rundle} D.,  {Harries} T.~J.,  {Acreman} D.~M.,    {Bate} M.~R.,  2010, MNRAS,
  407, 986

\bibitem[\protect\citeauthoryear{{Shu}}{{Shu}}{1977}]{shu_1977}
{Shu} F.~H.,  1977, ApJ, 214, 488

\bibitem[\protect\citeauthoryear{{Szebehely} \& {Peters}}{{Szebehely} \&
  {Peters}}{1967}]{szebehely_1967}
{Szebehely} V.,  {Peters} C.~F.,  1967, AJ, 72, 876

\bibitem[\protect\citeauthoryear{{Tan}, {Beltran}, {Caselli}, {Fontani},
  {Fuente}, {Krumholz}, {McKee} \& {Stolte}}{{Tan} et~al.}{2014}]{tan_2014}
{Tan} J.~C.,  {Beltran} M.~T.,  {Caselli} P.,  {Fontani} F.,  {Fuente} A.,
  {Krumholz} M.~R.,  {McKee} C.~F.,    {Stolte} A.,  2014, ArXiv e-prints

\bibitem[\protect\citeauthoryear{{Wolfire} \& {Cassinelli}}{{Wolfire} \&
  {Cassinelli}}{1987}]{wolfire_1987}
{Wolfire} M.~G.,  {Cassinelli} J.~P.,  1987, ApJ, 319, 850

\bibitem[\protect\citeauthoryear{{Yorke} \& {Sonnhalter}}{{Yorke} \&
  {Sonnhalter}}{2002}]{yorke_2002}
{Yorke} H.~W.,  {Sonnhalter} C.,  2002, ApJ, 569, 846

\end{thebibliography}
\end{document}